\documentclass[apjl]{emulateapj}

\pdfoutput=1

\usepackage{comment}
\usepackage{ifthen}
\usepackage{amsmath}
\usepackage{amsfonts}
\usepackage{amssymb}
\usepackage{wasysym}
\usepackage{subfigure}
\usepackage{enumitem}
\usepackage{wasysym}
\usepackage{tablefootnote}
\usepackage[%
	colorlinks=true,	
	urlcolor=blue,		
	pdfpagelabels=true,
	hypertexnames=true,
	plainpages=false,
	naturalnames=true,
	pdftitle={Priors for Orbital Eccentricity},    
	pdfauthor={David M. Kipping},     
	linkcolor=red,          
	citecolor=green,        
        ]{hyperref}
\newcommand{\luna}{{\tt LUNA}}
\newcommand{\multi}{{\sc MultiNest}}
\newcommand{\cofiam}{{\tt CoFiAM}}
\newcommand{\SNR}{\mathrm{SNR}}
\newcommand{\kepler}{\textit{Kepler}}

\newboolean{emulateapj}
\setboolean{emulateapj}{true}

\newboolean{astroph}
\setboolean{astroph}{true}

\shortauthors{Kipping et al.}
\shorttitle{V. A Survey of 41 KOIs for Exomoons}
\ifthenelse{\boolean{emulateapj}}{
    \newcommand{\titledag}{$\dagger$}
}{
    \newcommand{\titledag}{\dagger}
}

\begin{document}

\title {The Hunt for Exomoons with Kepler (HEK):\\
 V. A Survey of 41 Planetary Candidates for Exomoons
\altaffilmark{\titledag}}

\author{
	{\bf D.~M.~Kipping\altaffilmark{1},
             A.~R.~Schmitt\altaffilmark{2},
             X.~Huang\altaffilmark{3},
             G.~Torres\altaffilmark{4},\\
             D.~Nesvorn\'y\altaffilmark{5},
             L.~A.~Buchhave\altaffilmark{4,6},
             J.~Hartman\altaffilmark{3},
             G.~\'A.~Bakos\altaffilmark{3,7,8}
	}
}

\altaffiltext{1}{Dept. of Astronomy, Columbia University, 550 W 120th St., 
New York, NY 10027, USA; email: dkipping@astro.columbia.edu}

\altaffiltext{2}{Citizen Science}

\altaffiltext{3}{Dept. of Astrophysical Sciences, Princeton University,
		Princeton, NJ 05844, USA}

\altaffiltext{4}{Harvard-Smithsonian Center for Astrophysics,
		Cambridge, MA 02138, USA; email: dkipping@astro.columbia.edu}

\altaffiltext{5}{Dept. of Space Studies, Southwest Research Institute, 
1050 Walnut St., Suite 300, Boulder, CO 80302, USA}

\altaffiltext{6}{Centre for Star and Planet Formation, Natural History Museum of 
                 Denmark, University of Copenhagen, DK-1350 Copenhagen, Denmark}

\altaffiltext{7}{Alfred P. Sloan Fellow}

\altaffiltext{8}{Packard Fellow}

\altaffiltext{$\dagger$}{
Based on archival data of the \emph{Kepler} telescope. 
}


\begin{abstract}

We present a survey of 41 \kepler\ Objects of Interest (KOIs) for exomoons using 
Bayesian photodynamics, more than tripling the number of KOIs surveyed with
this technique. We find no compelling evidence for exomoons although thirteen 
KOIs yield spurious detections driven by instrumental artifacts, stellar 
activity and/or perturbations from unseen bodies. Regarding the latter, we find 
seven KOIs exhibiting $>5$\,$\sigma$ evidence of transit timing variations, 
including the ``mega-Earth'' Kepler-10c, likely indicating an additional 
planet in that system. We exploit the moderately large sample of 57 unique KOIs
surveyed to date to infer several useful statistics. For example, although there 
is a diverse range in sensitivities, we find that we are sensitive to 
Pluto-Charon mass-ratio systems for $\simeq40$\% of KOIs studied and Earth-Moon 
mass-ratios for 1 in 8 cases. In terms of absolute mass, our limits probe down
to 1.7 Ganymede masses, with a sensitivity to Earth-mass moons for 1 in 3 cases 
studied and to the smallest moons capable of sustaining an Earth-like atmosphere 
(0.3\,$M_{\oplus}$) for 1 in 4. Despite the lack of positive detections to date, 
we caution against drawing conclusions yet, since our most interesting objects 
remain under analysis. Finally, we point out that had we searched for the 
photometric transit signals of exomoons alone, rather than using photodynamics, 
we estimate that 1 in 4 KOIs would have erroneously been concluded to harbor 
exomoons due to residual time correlated noise in the \kepler\ data, posing a 
serious problem for alternative methods.

\end{abstract}



\section{INTRODUCTION}
\label{sec:intro}

The occurrence rate, and even the existence, of moons orbiting extrasolar 
planets is presently unknown. Their discovery would provide, for the first time, 
information on the physical processes governing satellite formation and
evolution outside the Solar System. Additionally, even null detections can guide 
theory in this respect, when carefully interpreted in terms of upper limits 
using the appropriate statistical techniques. Since it is a-priori unknown 
whether exomoons are sufficiently large to be detected using current 
instrumentation, this realization motivates a survey program to be designed such 
that statistically rigorous upper limits are derived for null detections. Whilst 
this strategy necessitates substantially greater computational resources, a 
survey adopting this philosophy has a guaranteed science product even if no 
confirmed objects are ever found, which in this case is the occurrence of 
exomoons as a function of the planet and satellite properties.

The ``Hunt for Exomoons with Kepler'' (HEK) project \citep{HEK1} is a survey 
for exomoons around transiting exoplanets which follows this ethos. Our survey 
uses Bayesian inference to derive rigorous upper limits in the cases of null
detections and employs a suite of vetting strategies to interrogate candidate
signals, honed by our past experience of conducting this survey. In previous
works, we surveyed 18 \kepler\ Objects of Interest (KOIs) for evidence of 
exomoons including a sample of M-dwarf host stars \citep{HEK4}, habitable-zone
KOIs \citep{HEK3}, dynamically active systems \citep{KOI872} and single KOIs 
\citep{HEK2}. In each case, we searched for both the transits of the exomoon
and the perturbations induced by its mass with a self-consistent, analytic,
dynamical model known as \luna\ \citep{LUNA}. Correlated noise structure is a 
particularly frequent source of exomoon transit false-positives, as outlined
in the cautionary tale of Kepler-90g.01 \citep{KEP90}.

In this paper, we dramatically increase the number of planetary candidates
surveyed by HEK, presenting an analysis of 41 KOIs. Since two of these were
already studied by previous surveys (but with fewer data), this paper brings the
total number of unique KOIs surveyed for exomoons by HEK to 57. The paper
is split as follows. In \S\ref{sec:targets}, we describe our target selection
strategy for these 41 KOIs via three distinct pathways. In \S\ref{sec:analysis}, 
the data handling, modeling and inference methods are described, including the 
vetting and classification strategies we employ. In \S\ref{sec:results}, we 
present the results of our survey, including refined planet parameters and an 
analysis of transit timing variations. We conclude in \S\ref{sec:discussion} 
with a discussion focussed on some statistical results from this sample and the 
HEK surveys to date, including empirical sensitivity limits and computational 
demands.

\section{TARGET SELECTION}
\label{sec:targets}

\subsection{Overview}

As outlined in \citet{HEK1}, the HEK project utilizes three complementary 
methods of selecting targets for analysis:

\begin{itemize}
\item[{\tiny$\blacksquare$}] \textbf{TSA:} Target Selection Automatic
\item[{\tiny$\blacksquare$}] \textbf{TSV:} Target Selection Visual
\item[{\tiny$\blacksquare$}] \textbf{TSO:} Target Selection via Opportunities
\end{itemize}

In previous papers, we limited our samples to targets selected using
just one of these methods. In this work, thanks to the larger number of KOIs
under consideration, we selected multiple targets from each category. 
Specifically, our sample consists of 16 KOIs selected via TSA, 16 via TSV and 9
via TSO, as described below. The full list of KOIs considered here is listed in 
Table~\ref{tab:targets}.

\begin{table*}
\caption{\emph{List of KOIs surveyed for exomoons in this work, and their basic 
parameters taken from the NASA Exoplanet Archive.
}} 
\centering 
\begin{tabular}{c c c c c c c c c c} 
\hline
KOI & $P_P$\,[d] & $R_P$\,[$R_{\oplus}$] & $\SNR$ & Multip. & 
$T_{\mathrm{eq}}$\,[K] & $T_{\mathrm{eff}}$\,[K] & $K\!p$ & Planet Name & Validation Ref. \\ [0.5ex] 
\hline
\multicolumn{2}{l}{\textbf{16 via TSA}} \\
\hline
KOI-0438.02 &  52.7 &  2.06 &  6.77 & 2 & 298 & 3985 & 14.3 & Kepler-155c & \citet{rowe:2014} \\
KOI-0518.03 & 247.4 &  2.14 &  9.74 & 3 & 212 & 4590 & 14.3 & Kepler-174d & \citet{rowe:2014} \\
KOI-0854.01 &  56.1 &  2.05 &  5.73 & 1 & 252 & 3593 & 15.8 & - & - \\
KOI-0868.01 & 236.0 &  8.73 &  7.13 & 1 & 172 & 3822 & 11.6 & - & - \\
KOI-1361.01 &  59.9 &  2.53 &  7.38 & 1 & 298 & 4014 & 15.0 & Kepler-61b & \citet{ballard:2013} \\
KOI-1431.01 & 345.2 &  8.00 &  4.65 & 1 & 272 & 5628 & 13.5 & - & - \\
KOI-1783.02 & 284.0 &  4.80 &  4.79 & 2 & 321 & 6215 & 13.9 & - & - \\
KOI-1830.02 & 198.7 &  2.88 &  6.03 & 2 & 257 & 5080 & 14.4 & - & - \\
KOI-1876.01 &  82.5 &  2.56 &  7.78 & 1 & 276 & 4237 & 15.2 & - & - \\
KOI-2020.01 & 111.0 &  2.32 & 10.17 & 1 & 260 & 4441 & 15.5 & - & - \\
KOI-2686.01 & 211.0 &  3.68 &  7.21 & 1 & 233 & 4628 & 13.9 & - & - \\
KOI-2691.01 &  97.5 &  3.30 &  6.35 & 1 & 313 & 4728 & 15.0 & - & - \\
KOI-3263.01 &  76.9 &  7.00 &  5.03 & 1 & 211 & 3587 & 15.9 & - & - \\
KOI-4005.01 & 178.1 &  2.22 &  5.55 & 1 & 316 & 5592 & 14.6 & Kepler-439b & \citet{torres:2015} \\
KOI-4036.01 & 168.8 &  2.24 &  5.93 & 1 & 279 & 4893 & 14.1 & - & - \\
KOI-4054.01 & 169.1 &  2.21 &  5.58 & 1 & 308 & 5380 & 14.6 & - & - \\
\hline
\multicolumn{2}{l}{\textbf{16 via TSV}} \\
\hline
KOI-0092.01 &  65.7 &  3.07 &  8.70 & 1 & 503 & 5952 & 11.7 & - & - \\
KOI-0112.01 &  51.1 &  3.04 &  6.74 & 2 & 552 & 5803 & 12.8 & - & - \\
KOI-0209.01 &  50.8 &  9.16 &  5.29 & 2 & 636 & 6466 & 14.3 & Kepler-117b & \citet{rowe:2014} \\
KOI-0276.01 &  41.7 &  2.71 &  5.86 & 1 & 629 & 5982 & 11.9 & - & - \\
KOI-0308.01 &  35.6 &  2.88 &  2.28 & 1 & 672 & 6210 & 12.4 & - & - \\
KOI-0374.01 & 172.7 &  3.14 &  9.26 & 1 & 389 & 5839 & 12.2 & - & - \\
KOI-0398.01 &  51.8 &  8.77 &  2.98 & 3 & 459 & 5271 & 15.3 & - & - \\
KOI-0458.01 &  53.7 &  8.16 &  2.65 & 1 & 558 & 5833 & 14.7 & - & - \\
KOI-0847.01 &  80.9 &  4.83 &  1.13 & 1 & 413 & 5665 & 15.2 & - & - \\
KOI-1535.01 &  70.7 &  2.13 &  7.67 & 1 & 539 & 6174 & 13.0 & - & - \\
KOI-1726.01 &  45.0 &  2.01 &  9.63 & 1 & 405 & 4684 & 13.1 & - & - \\
KOI-1808.01 &  89.2 &  3.82 &  5.36 & 1 & 504 & 6277 & 12.5 & - & - \\
KOI-1871.01 &  92.7 &  2.25 &  7.34 & 2 & 295 & 4528 & 14.9 & - & - \\
KOI-2065.01 &  80.2 &  3.60 &  4.33 & 1 & 437 & 5685 & 14.3 & - & - \\
KOI-2762.01 & 133.0 &  2.15 &  4.69 & 1 & 266 & 4530 & 15.0 & - & - \\
KOI-5284.01 & 389.3 &  6.79 &  2.10 & 1 & 267 & 5731 & 14.6 & - & - \\
\hline
\multicolumn{2}{l}{\textbf{9 via TSO}} \\
\hline
KOI-0072.02 &  45.3 &  2.30 & 12.44 & 2 & 554 & 5627 & 11.0 & Kepler-10c & \citet{dumusque:2014} \\
KOI-0245.01 &  39.8 &  1.88 & 49.27 & 3 & 456 & 5417 &  9.7 & Kepler-37d & \citet{barclay:2013} \\
KOI-0386.02 &  76.7 &  2.89 &  4.15 & 2 & 512 & 6197 & 13.8 & Kepler-146c & \citet{rowe:2014} \\
KOI-0518.02 &  44.0 &  1.43 & 10.37 & 3 & 377 & 4590 & 14.3 & Kepler-174c & \citet{rowe:2014} \\
KOI-0722.01 &  46.4 &  2.85 &  3.12 & 1 & 635 & 6343 & 13.5 & - & - \\
KOI-1783.01 & 134.5 &  7.93 &  6.06 & 2 & 412 & 6215 & 13.9 & - & - \\
KOI-2358.01 &  56.5 &  2.04 &  6.39 & 1 & 583 & 6412 & 13.5 & - & - \\
KOI-3663.01 & 282.5 & 11.28 &  5.20 & 1 & 328 & 6007 & 12.6 & PH-2b & \citet{wang:2013} \\
KOI-3681.01 & 217.8 & 22.02 &  2.58 & 2 & 497 & 6382 & 11.7 & - & - \\ [1ex]
\hline
\end{tabular}
\label{tab:targets} 
\end{table*}

\subsection{Target Selection Automatic (TSA)}
\label{sub:TSA}

The TSA algorithm is unchanged from that described in \citet{HEK4} and 
references therein. In summary, the algorithm assigns an estimated mass and
tidal properties (specifically $k_{2P}/Q$) to each KOI based on the size from
empirical and theoretical relations. Assuming loss through tidal evolution, we 
are then able to compute the maximum moon mass that can survive for 5\,Gyr 
using the expressions of \citet{barnes:2002}. This maximum moon mass is then
converted into a radius using an appropriate mass-radius relation and finally
into a signal-to-noise ratio ($\SNR$) using the \kepler\ Combined 
Differential Photometric Precision (CDPP) statistics \citep{christiansen:2012}.

Target KOIs were selected from the NASA Exoplanet Archive list of KOIs,
excluding objects flagged as false-positives. In this work, we elect to restrict 
the TSA targets to those KOIs with radii $R_P>2$\,$R_{\oplus}$, insolation
$S_{\mathrm{eff}}<6.925$\,$S_{\oplus}$ (inner edge of the habitable-zone defined
by \citealt{zsom:2013}) and $\SNR>4/\sqrt{0.75}=4.62$. The $\sqrt{0.75}$ scaling 
is based on the fact TSA computes $\SNR$ assuming one uses the entire time 
series, but in reality we only use 75\% of the data in our fits for reasons 
described in \S\ref{sec:analysis}.

In total, 25 targets satisfied these criteria, of which three have already been 
studied in previous surveys (Kepler-22b, \citealt{HEK3}; KOI-1596.02, 
\citealt{HEK4}; KOI-1876.01, \citealt{HEK2}). Since significantly more data is
now available for KOI-1876.01 than available for the analysis in \citet{HEK2}, 
this target was kept in our TSA sample. KOI-1871.01 and KOI-2762.01 were
removed from this subset because they were flagged as candidates by the TSV 
process described later in \S\ref{sub:TSV}. Additionally, KOI-1174.01, 
KOI-490.02 and KOI-1274.01 (also known as Kepler-421b; \citealt{KEP421}) have 
fewer than three visible transits and were rejected. Finally, KOI-902.01 was 
reported by \citet{mazeh:2013} to display high-amplitude ($\gtrsim$100\,mins) 
and long-period ($\gtrsim3$\,years) transit timing variations (TTVs), which are 
uncharacteristic of exomoons \citep{kipping:2009}, and so was also rejected. 
This resulted in the 16 KOIs listed in Table~\ref{tab:targets}.

\subsection{Target Selection Visual (TSV)}
\label{sub:TSV}

In previous HEK papers, the only target picked for analysis using our TSV 
process was KOI-0872.01, also known as Kepler-46b (see the Supporting Online 
Material of \citealt{KOI872}). This target displayed in-transit flux increases 
characteristic of star-planet-moon mutual events (as well as large TTVs), but 
these were ultimately attributed to star-spot crossings (e.g. see 
\citealt{beky:2014}) rather than due to an exomoon.

In the earlier work, TSV was conducted using several publicly available light 
curve viewers. However, we quickly found that these viewers did not meet our 
needs for performing large scale visual surveys. In a typical survey, four to
five thousand KOI transit signals need to be examined for exomoon signals
across several hundred distinct planetary candidates. To effectively handle such
a large volume of data in a reasonable time frame requires a system with
advanced navigation and signal processing features, which the public tools did
not provide.

\begin{figure*}
\begin{center}
\includegraphics[width=19.0 cm,angle=0,clip=true]{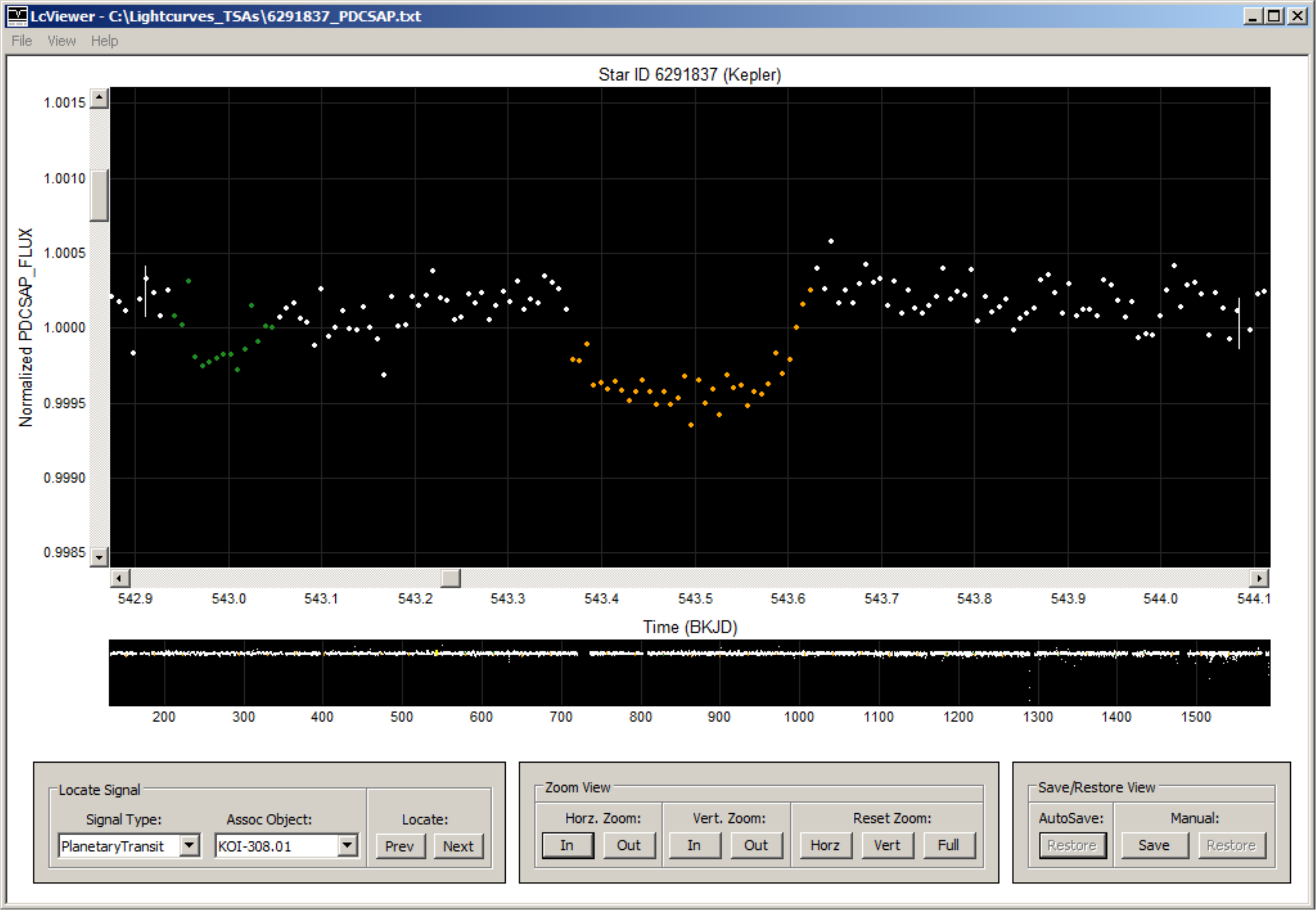}
\caption{
Main application window for LcViewer. Two signals are shown in this example: 
a planetary transit signal of KOI-0308.01 highlighted in orange and a user 
defined moon-like signal highlighted in green. The white vertical lines on 
either side of the light curve indicate the region of visual inspection, which 
is equal to the estimated Hill timescale.
} 
\label{fig:TSV}
\end{center}
\end{figure*}

To address these needs, we developed a custom designed software package known as 
LcTools. This package consists of two main components: LcGenerator for 
generating normalized light curve files in bulk, and LcViewer for examining light 
curve files and recording signals. This Windows-based software is freely 
available and can be obtained from the author (A. Schmitt) upon request. We 
direct those interested to the online description for technical 
details\footnote{See http://www.exomoon.org for details}, but we briefly point 
out some relevant features here:

\begin{itemize}[leftmargin=*]
\item[{\tiny$\blacksquare$}]
Bin short-cadence (SC) data to a user specified cadence.
\item[{\tiny$\blacksquare$}]
Independently pan and scale the time and flux axes.
\item[{\tiny$\blacksquare$}]
Locate each instance of a periodic signal and automatically scale the signal for 
optimized viewing.
\item[{\tiny$\blacksquare$}]
Automatically highlight known transit signals according to the NASA Exoplanet
Archive ephemerides (e.g. see Figure~\ref{fig:TSV}).
\item[{\tiny$\blacksquare$}]
Automatically highlight mutual planetary transit signals (the simultaneous 
transit of two or more planets).
\item[{\tiny$\blacksquare$}]
Record and highlight user-defined signals (e.g. due to exomoons), as
shown in Figure~\ref{fig:TSV}.
\end{itemize}

In total, 1289 unique KOIs were examined for exomoon signals across numerous 
surveys. Many KOIs were inspected multiple times. KOIs included both TSAs and 
non-TSAs. The latter group was largely composed of promising targets meeting the 
selection criteria below:

\begin{itemize}[leftmargin=*]
\item[{$\RHD$}]
Semi-major axis; $a \geq 0.1$\,AU (for reasons discussed in \citealt{namouni:2010}).
\item[{$\RHD$}]
Planet size; $R_P \geq 3$\,$R_{\oplus}$.
\item[{$\RHD$}]
Expected $\SNR$ for an exomoon satisfies $\SNR \geq 1$.
\end{itemize}

In all cases, the minimum transit duration required was four hours. From
experience, we find that at least eight long-cadence (LC) data points are
necessary to reliably detect exomoon signals visually.

For a given host star, all available Presearch Data Conditioning (PDCSAP) data 
(\citealt{stumpe:2012}, \citealt{smith:2012}) were obtained from the Barbara A. 
Mikulski Archive for Space Telescopes\footnote{https://archive.stsci.edu/kepler} 
(MAST). The data were normalized and merged into a single light curve file. Both 
the LC and SC data were used, with SC given priority. To mitigate the flux 
scatter and improve signal detection, SC data were binned to a lower data 
rate - typically 6, 10, or 12 data points per hour.

For each KOI transit signal, a visual search was made for moon-like signals 
within the marked KOI signal region (see Figure~\ref{fig:TSV} for an example). 
The KOI signal region is defined as the extended Hill timescale per the equation 
given in \S2.2 of \citet{HEK3}. Upon finding a moon-like signal conforming to 
the general shape characteristics predicted by \luna\ \citep{LUNA}, the signal 
was manually recorded in LcViewer and a screen snapshot taken.

A total of 2928 moon-like signals were recorded; 923 from the TSA 
group and 2005 from the non-TSA group. From this set, the highest quality 
signals were visually identified. For the current paper, 16 KOIs were found 
having at least three high quality exomoon signals (see 
Table~\ref{tab:targets}). Note that this represents just a fraction of the 
KOIs with high-quality signals found.

\subsection{Target Selection via Opportunities (TSO)}
\label{sub:TSO}

In addition to the 32 KOIs selected using the TSA and TSV methods, we picked
9 KOIs as ``targets of opportunity'' via TSO. This process is loosely defined
as identifying systems for which we have reason to suspect an elevated 
probability of a successful detection or that the system has some unusual
properties making an analysis of greater scientific value. As an example from
previous surveys, in \citet{HEK3} we selected Kepler-22b using TSO, since it
was claimed to be the first habitable-zone transiting planet discovered
\citep{borucki:2012}. Below, we briefly describe the reasons why each KOI was 
selected using TSO.

KOI-0722.01 was previously surveyed in \citet{HEK2} using quarters Q1-9 of
the \kepler\ data, but is re-analyzed here using Q1-17 due to the unusual fits
found previously.

KOI-0072.01, also known as Kepler-10c, is a confirmed transiting planet with a
mass measured using radial velocities \citep{dumusque:2014}. Argued by the
authors to be a solid Neptune-mass planet, Kepler-10c may represent a scaled
up version of the Earth leading us to wonder if there is a scaled up version of
the Moon in orbit too.

KOI-3663.01, also known as PH-2b, is a validated transiting planet 
\citep{wang:2013} found by the Planet Hunters project \citep{fischer:2012}.
The discoverers argue that the planet is a Jupiter-sized planet in the 
habitable-zone of its host star, making it an ideal target for an exomoon search.

KOI-1783.01 and KOI-0518.02 (also known as Kepler-174c; \citealt{rowe:2014})
are additional KOIs to two targets already selected in the TSA part of this 
work, namely KOI-1783.02 and KOI-0518.03 (also known as Kepler-174d; 
\citealt{rowe:2014}). Moreover, both of these have relatively high
$\SNR$ expected due to putative exomoons and so were selected via
TSO.

KOI-0245.01, also known as Kepler-37d, is a validated transiting planet and
the largest and widest orbit planet in a three-planet system 
\citep{barclay:2013}. The photometric precision achieved for Kepler-37 is 
exceptional, thanks to the star's brightness (9.7 in \kepler's bandpass) and low 
intrinsic noise, allowing for the discovery of the Moon-sized planet Kepler-37b.
For this reason, we added this target through TSO.

KOI-3681.01 was surveyed for exomoons in an independent study 
\citep{lewis:2014}, but only using the transit method rather than the full 
photodynamic modeling employed by HEK. Nevertheless, we added this target to
exploit the unique opportunity to compare our results.

Finally, KOI-0386.02 and KOI-2358.01 were selected as somewhat ``wildcard''
targets due to their moderate orbital periods providing plenty of transits
($76.7$\,d and $56.5$\,d respectively) and yet TSA still predicting reasonably
good $\SNR$ values of 4.15 and 6.39 respectively.

\section{ANALYSIS METHODOLOGY}
\label{sec:analysis}

\subsection{Data Handling}
\label{sub:data}

Photometric time series for the 41 KOIs studied in this work were obtained from
the MAST archive. We downloaded all available quarters from Q0-17 for each KOI 
and preferentially use the SC over LC data, where available. Only data with SAP 
QUALITY values equal to zero were utilized in this work, thereby rejecting data 
likely corrupted by known effects. 

As with previous papers, we only use the Simple Aperture Photometry (SAP) 
processed data in our analysis, rather than the Presearch Data Conditioning 
(PDCSAP) time series (\citealt{stumpe:2012}, \citealt{smith:2012}). The
PDCSAP data is optimized to remove common mode instrumental effects in order
to retain stellar astrophysics such as rotational modulations. However, in our
work, retaining the stellar astrophysics is not required and critically the act
of cleaning common modes will remove frequencies across the Fourier domain that
may potentially distort, remove or even spuriously introduce the very low 
amplitude exomoon-like signals we seek here.

Transit light curves are detrended on an epoch-by-epoch basis. Transits are
initially located using the NASA Exoplanet Archive reported ephemeris and
we consider a window of $\pm0.5P$ around each event. Discontinuities in the time 
series, due to pointing tweaks for example, are visually identified and in such
cases our epoch window is shortened up to the nearest such event surrounding
each transit i.e. we make no attempt to correct such effects. Long-term trends,
due to focus drift for example, are removed using the \cofiam\ algorithm
described in \citet{HEK2,HEK4}. This algorithm is essentially a low-cut,
high-pass filter removing all long-term trends with a threshold frequency set to
three times the NASA Exoplanet Archive reported transit duration. As described
in our previous papers, \cofiam\ ensures the transit light curve of the planet 
and any moons are not distorted, whilst removing long-term trends and minimizing 
the autocorrelation on a 30\,minute timescale. It is important to stress that
\cofiam\ does not remove high frequency noise, but our philosophy here is that
whilst one can consider alternative methods to accomplish this feat, it is 
unclear whether such methods can retain the guarantee of \cofiam\ that the 
transit shape is undisturbed.

\subsection{Forward Modeling}
\label{sub:luna}

As with previous HEK papers, the forward model that is compared to the
observations is the \luna\ algorithm, described in \citet{LUNA}. At every time
stamp, \luna\ calculates the sky-projected positions of the planet and moon
relative to the star, using a nested Keplerian body appropriate for a 
hierarchical three-body problem (see Appendix of \citealt{weighing:2010} for 
proof). Based on these positions, the amount of star light blocked by the planet 
and the moon are calculated analytically using the \citet{fewell:2006} and 
\citet{LUNA} solutions. Limb darkening of the star is appropriately accounted 
for, provided the radius of the satellite is small relative to the star i.e. 
$R_S\ll R_{\star}$ (this is not required for the planet). In this work, we use 
the quadratic limb darkening formulation described in \citet{LDfitting:2013} and 
account for long-cadence smearing using the re-sampling treatment described in 
\citet{binning:2010} ($N_{\mathrm{resam}}=30$). Quarter-to-quarter contaminating 
light factors are taken from the fits file (``CROWDSAP'' header keyword) and 
accounted for following the method described in \citet{nightside:2010}.

Recently, there have been several proposals for finding exomoons using
alternative methods to photodynamics, such as excess noise around transits
(``scatter-peak''; \citealt{simon:2012}), high frequency TTVs \citep{szabo:2013} 
and phase-stacking transits (``OSE''; \citealt{OSE:2014}), and we briefly
comment on why we still favor the photodynamical technique.

The alternative methods aim to be computationally cheaper than photodynamical 
modeling, at the expense of approximating the complex motions involved. However, 
it is important to note that photodynamical modeling accounts for all known 
photometric effects caused by a transiting exomoon, e.g., TTV 
\citep{sartoretti:1999}, TDV-V \citep{kipping:2009}, TDV-TIP 
\citep{kipping:2009b}, plus even previously unknown effects such ingress/egress
asymmetries \citep{LUNA}. Therefore, these simpler methods can all be understood 
to be approximations to the full photodynamical description and by definition 
model less of the information content imparted into the light curve by the 
presence of exomoon(s). With less information, they are all guaranteed to not 
only be less sensitive in terms of signal-to-noise, but also critically lacking
the powerful vetting techniques available with photodynamics, such as dynamical
weighing of the parent planet \citep{weighing:2010}. We also stress that 
our implementation of photodynamics models the entire \kepler\ time series, i.e. 
every transit is modeled simultaneously and self-consistently. Naturally then, 
there is no more signal-to-noise to be gained by phase-folding or binning 
techniques (e.g. \citealt{simon:2012}; \citealt{OSE:2014}), which will actually 
only serve to further degrade the information content. For these reasons, we 
argue that photodynamics remains the most robust method of searching for 
exomoons.

However, the alternative methods are typically much faster and may be useful
for identifying interesting systems worthy of more detailed analysis with
photodynamics i.e. as a prioritization tool. The HEK project currently adopts
this position in its application of the Target Selection Visual process
discussed in \S\ref{sub:TSV}.

\subsection{Bayesian Inference \& Model Selection}
\label{sub:fits}

For reasons discussed in \S\ref{sec:intro}, the HEK project places great 
importance on the use of rigorous statistical inference techniques in our 
survey. There are several unique needs of an exomoon survey that shape our 
choice of methodology:

\begin{itemize}
\item[{\tiny$\blacksquare$}]
The parameter space is complex, highly multimodal and features curved 
degeneracies, posing a major challenge to many conventional regression 
techniques.
\item[{\tiny$\blacksquare$}]
In order to derive an upper limit on the satellite-to-planet mass ratio, 
$(M_S/M_P)$, one must marginalize over the other parameters, such as the moon's
semi-major axis and the three-dimensional architecture of the orbits.
\item[{\tiny$\blacksquare$}]
A planet+moon model has significantly greater freedom than a planet-only case 
guaranteeing a higher maximum likelihood, thereby necessitating the use of
Bayesian model selection.
\end{itemize}

It should be recognized that satisfying these needs cannot be accomplished 
without high computational cost. One of the most efficient algorithms that
satisfies these requirements is \multi\ \citep{feroz:2008,feroz:2009}, based
on nested sampling \citep{skilling:2004}, and this is the method employed in 
this work and in our previous HEK papers. Using \multi, multiple objects (in
our case multiple exomoons) are identifiable as separate modes 
\citep{feroz:2008,feroz:2013}, unless the moons strongly interact. This
important feature, utilized since the inaugural HEK paper \citep{HEK1}, 
refutes the claim made in \citet{OSE:2014} that OSE is the first method 
permitting the characterization of multiple moon systems.

For each KOI, we regress at least two models, or hypotheses, to the data. The 
null hypothesis is $\mathcal{H}_P$, a planet-only case (forward model is that
of \citealt{mandel:2002}), which is compared to the planet+satellite hypothesis, 
$\mathcal{H}_S$ (forward model is \luna). The parameterization of these models 
is the same as that described in \citet{HEK4} and the priors adopted are listed 
in Table~\ref{tab:priors}. For both models, a Gaussian likelihood function is
adopted in our fits and exomoons are assumed to have near-circular orbits.

As in our previous paper \citet{HEK4}, we also have latent priors on the 
derived mean planetary and satellite densities, $\rho_P$ and $\rho_S$ (see 
\citealt{weighing:2010} for how these may be calculated). Specifically, we 
impose a maximum density for both of 27.9\,g\,cm$^{-3}$ (see \citealt{HEK1} for 
justification) and a minimum density for $\rho_P$ of 0.08\,g\,cm$^{-3}$ (no such 
minimum is imposed on $\rho_S$ to allow null detections of $M_S/M_P$). The 
purpose of these latent priors is to restrict \multi\ to only the physically 
plausible range of solutions (thereby imposing detection criteria B3 
automatically; see \S\ref{sub:vetting}). One drawback of this decision is that a 
zero-radius moon (i.e. a null detection of a moon signal) is forbidden, since 
$\rho_S\to\infty$. In future surveys, we will likely remove these latent priors 
to allow us to define upper limits on $(R_S/R_P)$ as well as the usual 
$(M_S/M_P)$.

Although the photometry has been detrended and normalized using \cofiam,
we allow the out-of-transit baseline flux to be varied in our fits too. This
term is linear with respect to the transit model and thus can be minimized 
analytically at each Monte Carlo realization, which we implement here following
the technique described in \citet{kundurthy:2011}.

\begin{table}
\caption{\emph{Priors adopted for the planet and moon parameters in our fits.
}} 
\centering 
\begin{tabular}{c c} 
\hline
Parameter & Prior$^{\dagger}$ \\ [0.5ex] 
\hline
\multicolumn{2}{l}{\textit{Planet Parameters}} \\
\hline
$p$ & $\mathcal{U}\{0,1\}$ \\
$\log_{10}(\rho_{\star}$\,[$\mathrm{g}$\,$\mathrm{cm}^{-3}])$ & $\mathcal{U}\{-3,3\}$ \\
$b$ & $\mathcal{U}\{0,2\}$ \\
$P$\,$[\mathrm{days}]$ & $\mathcal{U}\{\hat{P}-1,\hat{P}+1\}^{\ast}$ \\
$\tau$\,[days] & $\mathcal{U}\{\hat{\tau}-1,\hat{\tau}+1\}^{\ast}$ \\
$q_1^{\times}$ & $\mathcal{U}\{0,1\}$ \\
$q_2^{\times}$ & $\mathcal{U}\{0,1\}$ \\
\hline
\multicolumn{2}{l}{\textit{Moon Parameters}} \\
\hline
$(R_S/R_P)$ & $\mathcal{U}\{-1,1\}$ \\
$(M_S/M_P)$ & $\mathcal{U}\{0,1\}$ \\
$a_{SP}$ & $\mathcal{U}\{2,7.897 \hat{P}^{2/3}\}^{\ast}$ \\
$\cos i_S'^{\diamond}$ & $\mathcal{U}\{-1,3\}$ \\
$\Omega_S$\,\,[rads] & $\mathcal{U}\{-\pi,\pi\}$ \\
$P_S$\,\,$[\mathrm{days}]$ & $\mathcal{J}\{0.052,\hat{P}/\sqrt{3}\}^{\ast}$ \\
$\phi_S$\,\,[rads] & $\mathcal{U}\{0,2\pi\}$ \\ [1ex]
\hline 
\multicolumn{2}{l}{$^{\dagger}$ $\mathcal{U}$ and $\mathcal{J}$ denote a uniform and Jeffrey's prior} \\
\multicolumn{2}{l}{$^{\ast}$ $\hat{P}$ and $\hat{\tau}$ are the reported ephemeris parameters} \\
\multicolumn{2}{l}{\,\,\,\,\,\,taken from the NASA Exoplanet Archive} \\
\multicolumn{2}{l}{$^{\times}$ Defined in \citet{LDfitting:2013}} \\
\multicolumn{2}{l}{$^{\diamond}$ Defined in \citet{HEK3}} \\
\multicolumn{2}{l}{\,} \\
\end{tabular}
\label{tab:priors} 
\end{table}

\subsection{Vetting}
\label{sub:vetting}

For each KOI studied, we obtain the Bayesian evidence and parameter posterior 
distributions for the planet-only model, $\mathcal{H}_P$, and also the 
planet-with-moon model, $\mathcal{H}_S$. This information must be used to
assess whether the KOI in question can be claimed to exhibit compelling evidence
for the presence of one or more exomoons. In \citet{HEK2}, we first defined a
list of standardized basic criteria that we use to make this assessment:

\begin{itemize}
\item[{\textbf{B1}}] Improved evidence of the planet-with-moon fits at 
$\geq 4$\,$\sigma$ confidence.
\item[{\textbf{B2}}] Planet-with-moon Bayesian evidences indicate both a mass 
and radius preference for the satellite.
\item[{\textbf{B3}}] Parameter posteriors are physical.
\item[{\textbf{B4}}] Mass and radius of the moon converge away from zero.
\end{itemize}

As discussed in \S\ref{sub:fits}, criterion B3 is enforced for all
fits by virtue of our latent priors. This forces the mean density of the planet
and the satellite to be within the extreme range discussed in \citet{HEK1}. It
is therefore not explicitly tested. Criterion B1 remains unchanged in
this work. Although Bayesian model selection is an important prerequisite for
claiming a detection, it is insufficient in isolation since we adopt a Gaussian
likelihood function to compute the evidences, and real photometry is never 
perfectly Gaussian.

As of \citet{HEK3}, criteria B2 was modified. Formally, B2 requires running two 
extra photodynamical models in addition to $\mathcal{H}_S$, which naturally 
slows down our survey speed considerably. In \citet{HEK3}, we proposed focussing 
on the radius aspect of B2 alone, which we dub B2a, and testing the condition by 
allowing $\mathcal{H}_S$ to explore ``negative radius'' exomoons. Such signals 
correspond to inverted transits and we therefore require a positive radius moon 
as part of our vetting tests. Negative-radius moons have no physical
meaning and are most likely indications of time-correlated noise structure.

For criterion B4, the radius test is not useful for vetting when a latent prior 
exists on $\rho_S$, such as done in this work, since zero-radius moons are 
forbidden. Instead, we focus on the mass component which remains unbiased and 
dub the criterion B4a.

These modifications lead to the following basic detection criteria:

\begin{itemize}
\item[{\textbf{B1}}] Improved evidence of the planet-with-moon fits at 
$\geq 4$\,$\sigma$ confidence
\item[{\textbf{B2a}}] Planet-with-moon evidences indicate a preference for
a positive radius moon (34.13\% quantile of $(R_S/R_P)$ is positive)
\item[{\textbf{B4a}}] Mass of the moon converges away from zero (false alarm
probability of a zero-mass moon is $<5$\% using the \citealt{lucy:1971} test).
\end{itemize}

Additionally, \citet{HEK2} defined several ``follow-up'' criteria of which the 
most powerful is F2: ``The predictive power of the moon model is superior
(or at least equivalent) to that of a planet-only model''. This test first
requires that we only fit $\sim75$\% of the transits in models $\mathcal{H}_P$
and $\mathcal{H}_S$. We then a) extrapolate the maximum a-posteriori realization
of the two hypotheses and require $\mathcal{H}_S$ to yield a better likelihood
and b) same as a), except we repeat for $10^4$ random draws from the posterior
distributions of the two hypotheses and require $\mathcal{H}_S$ to be superior
in 95\% of the realizations. These respectively define criteria F2a and F2b.

In practice, we only conduct F2b if F2a is first passed, to save resources. In
some instances, the limited number of transits available meant we elected to fit
100\% of the data, making F2 unavailable as a test. Otherwise, the $\sim25$\% 
of ignored data is chosen to be near the median of the available LC and SC 
transits.

F2 is a powerful tool in vetting exomoons but will be passed by KOIs exhibiting
high signal-to-noise sinusoidal TTVs, since a moon-like TTV is also a sinusoid
\citep{kipping:2009}. KOI-0308.01 and KOI-2691.01 both show very strong evidence 
for TTVs ($>17$\,$\sigma$) and nearly sinusoidal-like behavior, meaning they 
both passed F2a and F2b. As a more useful test in these two instances, we define
the null model to be a finite-mass, zero-radius moon model $\mathcal{H}_{S,R0}$.
This model is able to reproduce the sinusoidal TTVs but clearly a real exomoon
case should be favored over a model negating the moon's radius.

\subsection{Classification}
\label{sub:classification}

Naively, one might expect a KOI to either be classified as a confirmed exomoon
or a null result. In practice, experience has revealed that spurious detections 
are common and are generally unsuitable for deriving upper limits. For this 
reason, it is necessary to classify null detections (``ND'') versus 
false-positives (``FP''). In our project, the upper limit we are interested in 
is the mass-ratio $(M_S/M_P)$, since $(R_S/R_P)$ is known to be both less 
sensitive and less reliable e.g. see \citet{HEK2,HEK3,KEP90}. Therefore,
we define null detections as being cases where the posterior distribution of 
$(M_S/M_P)$ does not converge away from zero. This is equivalent to criterion 
B4a, providing a quantitative method of classifying null detections.

False-positives can be due to a variety of effects and in many cases we are
unable to find an unambiguous explanation (except that we know an exomoon is 
unlikely to be responsible). These unknown false positives are dubbed ``FP-U''. 
In cases where instrumental effects are thought to be responsible, we dub the 
case ``FP-I''. A recent example of this was discussed in detail for 
Kepler-90g.01 in \citet{KEP90}, where a pixel level effect was found to 
introduce an exomoon false-positive. Finally, if perturbations from another body 
are likely responsible, we dub the case ``FP-P''. These are identified by 
assessing whether the TTVs show a significant signal, with $\geq4$\,$\sigma$ 
confidence set as our threshold for classification as FP-P.

\\

\section{RESULTS}
\label{sec:results}

\subsection{Exomoon Survey Results}
\label{sub:exomoonresults}

\subsubsection{Overview}

Out of the 41 KOIs surveyed in this work, we find no compelling evidence for
exomoons. Using our detection criteria discussed in \S\ref{sub:vetting}, we find 
that only KOI-0072.01 (Kepler-10c) and KOI-3663.01 (PH-2b) pass these 
standardized tests. However, in both cases we conclude that the moon candidates 
are false-positives, albeit for differing reasons discussed shortly.

\begin{table*}
\caption{\emph{Summary of survey results. KOIs with multiple rows denote
multiple modes.
}} 
\centering 
\begin{tabular}{c c c c c c c c c c} 
\hline
KOI & $(M_S/M_P)^{\ast}$ & $(R_S/R_P)$ & B1 & B2a & B4a & F2a & F2b & Classification & Est. $M_S$ [$M_{\oplus}]$ \\ [0.5ex] 
\hline
KOI-0438.02 & $<0.96$ & $-0.735_{-0.027}^{+0.028}$ &
\checkmark & 
\text{\sffamily X} &  
\text{\sffamily X} &
N/A & N/A & ND & $<5.07$ \\ 
KOI-0518.03 & $<0.91$ & $+0.48_{-0.31}^{+0.27}$ &
\checkmark & 
\checkmark & 
\text{\sffamily X} &
N/A & N/A & ND & $<5.51$ \\
KOI-0854.01 & $<0.94$ & $+0.78_{-1.23}^{+0.15}$ &
\text{\sffamily X} & 
\checkmark & 
\text{\sffamily X} & 
\text{\sffamily X} & 
- & ND & $<5.33$ \\ 
KOI-0868.01 & $0.0142_{-0.0010}^{+0.0012}$ & $-0.076_{-0.012}^{+0.014}$ & 
\text{\sffamily X} & 
\text{\sffamily X} &
\checkmark & 
N/A & N/A & FP-P & - \\
KOI-1361.01 & $<0.87$ & $+0.66_{-0.18}^{+0.18}$ &
\text{\sffamily X} & 
\checkmark & 
\text{\sffamily X} &
\text{\sffamily X} &
- & ND & $<5.36$ \\ 
KOI-1431.01 & $0.00171_{-0.00023}^{+0.00024}$ & $-0.141_{-0.014}^{+0.015}$ & 
\checkmark & 
\text{\sffamily X} &
\checkmark &
N/A & N/A & FP-P & - \\
KOI-1783.02 & $0.49_{-0.16}^{+0.24}$ & $-0.264_{-0.037}^{+0.037}$ &
\checkmark & 
\text{\sffamily X} & 
\checkmark &   
N/A & N/A & FP-P & - \\ 
KOI-1830.02 & $<0.15$ & $+0.445_{-0.089}^{+0.085}$ &
\text{\sffamily X} &
\checkmark & 
\text{\sffamily X} &   
\checkmark & 4
\text{\sffamily X} & 
ND & $<1.14$ \\ %
KOI-1876.01 & $<0.047$ & $-0.547_{-0.052}^{+0.057}$ &
\text{\sffamily X} & 
\text{\sffamily X} & 
\text{\sffamily X} & 
N/A & N/A & ND & $<0.31$ \\
KOI-2020.01 & $<0.0019$ & $-0.540_{-0.058}^{+0.055}$ &
\checkmark & 
\text{\sffamily X} & 
\text{\sffamily X} &  
\text{\sffamily X} & 
- & ND & $<0.11$ \\ 
KOI-2686.01 & $0.153_{-0.019}^{+0.028}$ & $-0.286_{-0.035}^{+0.041}$ &
\checkmark & 
\text{\sffamily X} & 
\checkmark & 
\text{\sffamily X} & 
- & FP-P & - \\ 
- & $0.68_{-0.18}^{+0.17}$ & $-0.266_{-0.046}^{+0.050}$ &
\checkmark & 
\text{\sffamily X} & 
\checkmark & 
\text{\sffamily X} & 
- & FP-P & - \\ 
KOI-2691.01$^{\dagger}$ & $0.63_{-0.20}^{+0.23}$ & $+0.182_{-0.040}^{+0.043}$ &
\text{\sffamily X} &  
\checkmark &
\checkmark & 
\text{\sffamily X} & 
- & 
FP-P & - \\ 
KOI-3263.01 & $<0.0020$ & $-0.073_{-0.018}^{+0.017}$ &
\text{\sffamily X} & 
\text{\sffamily X} & 
\text{\sffamily X} &  
N/A & N/A & ND & $<0.041$ \\
KOI-4005.01 & $<0.30$ & $-0.471_{-0.079}^{+0.094}$ &
\text{\sffamily X} & 
\text{\sffamily X} & 
\text{\sffamily X} & 
\checkmark & 
\text{\sffamily X} & 
ND & $<1.91$ \\ 
KOI-4036.01 & $<0.048$ & $+0.581_{-0.115}^{+0.098}$ &
\text{\sffamily X} & 
\checkmark &
\text{\sffamily X} &  
\text{\sffamily X} & 
- & 
ND & $<0.26$ \\
KOI-4054.01 & $<0.069$ & $-0.713_{-0.086}^{+0.085}$ &
\text{\sffamily X} & 
\text{\sffamily X} & 
\text{\sffamily X} &  
\text{\sffamily X} & 
- & ND & $<0.40$ \\ 
\hline
KOI-0092.01 & $0.168_{-0.082}^{+0.075}$ & $+0.307_{-0.066}^{+0.066}$ &
\text{\sffamily X} &
\checkmark & 
\checkmark & 
\text{\sffamily X} & 
- & FP-U & - \\ 
- & $<0.017$ & $-0.217_{-0.048}^{+0.037}$ &
\text{\sffamily X} & 
\text{\sffamily X} & 
\text{\sffamily X} &  
\text{\sffamily X} & 
- & ND & $<0.11$ \\ 
KOI-0112.01 & $<0.26$ & $+0.306_{-0.044}^{+0.049}$ &
\text{\sffamily X} & 
\checkmark & 
\text{\sffamily X} & 
\checkmark & 
\text{\sffamily X} & 
ND & $<2.18$ \\
KOI-0209.01 & $<0.010$ & $-0.3003_{-0.0072}^{+0.0075}$ &
\checkmark &  
\text{\sffamily X} & 
\text{\sffamily X} &   
\text{\sffamily X} & 
- & ND & $<1.12$ \\ 
KOI-0276.01 & $<0.65$ & $-0.26_{-0.11}^{+0.64}$ &
\text{\sffamily X} & 
\text{\sffamily X} & 
\text{\sffamily X} &  
\text{\sffamily X} & 
- & ND & $<4.31$ \\
KOI-0308.01$^{\dagger}$ & $0.952_{-0.070}^{+0.036}$ & $-0.338_{-0.019}^{+0.020}$ &
\text{\sffamily X} & 
\text{\sffamily X} & 
\checkmark &
\checkmark & 
\text{\sffamily X} &
FP-P & - \\ 
KOI-0374.01 & $<0.60$ & $+0.205_{-0.039}^{+0.050}$ &
\checkmark & 
\checkmark & 
\text{\sffamily X} & 
\text{\sffamily X} &  
- & ND & $<5.03$ \\
KOI-0398.01 & $<0.021$ & $-0.181_{-0.038}^{+0.041}$ &
\text{\sffamily X} &  
\text{\sffamily X} &
\text{\sffamily X} & 
\text{\sffamily X} & 
- & ND & $<1.47$ \\ 
KOI-0458.01 & $0.74_{-0.20}^{+0.15}$ & $+0.857_{-0.120}^{+0.087}$ &
\text{\sffamily X} & 
\checkmark & 
\checkmark &  
\text{\sffamily X} & 
- & FP-U & - \\ 
- & $<0.88$ & $+0.66_{-0.23}^{+0.21}$ &
\text{\sffamily X} & 
\checkmark & 
\text{\sffamily X} &
\text{\sffamily X} & 
- & ND & $<71.6$ \\ 
KOI-0847.01 & $<0.024$ & $-0.278_{-0.025}^{+0.023}$ &
\checkmark &
\text{\sffamily X} &  
\text{\sffamily X} & 
\text{\sffamily X} & 
- & ND & $<0.062$ \\
KOI-1535.01 & $<0.012$ & $-0.469_{-0.041}^{+0.049}$ &
\text{\sffamily X} &
\text{\sffamily X} & 
\text{\sffamily X} &  
\text{\sffamily X} &
- & ND & $<0.077$ \\ 
KOI-1726.01 & $<0.17$ & $-0.412_{-0.032}^{+0.038}$ &
\text{\sffamily X} & 
\text{\sffamily X} & 
\text{\sffamily X} & 
\text{\sffamily X} & 
- & ND & $<0.97$ \\ 

KOI-1808.01 & $0.35_{-0.11}^{+0.17}$ & $+0.765_{-0.085}^{+0.110}$ &
\checkmark & 
\checkmark &  
\checkmark & 
\text{\sffamily X} & 
- & FP-U & - \\
KOI-1871.01 & $<0.92$ & $-0.550_{-0.066}^{+0.083}$ &
\text{\sffamily X} & 
\text{\sffamily X} &  
\text{\sffamily X} &  
\checkmark & 
\text{\sffamily X} & 
ND & $<5.36$ \\ 
KOI-2065.01 & $<0.40$ & $+0.669_{-0.070}^{+0.077}$ &
\text{\sffamily X} & 
\checkmark & 
\text{\sffamily X} &   
\text{\sffamily X} & 
- & ND & $<2.89$ \\ 
KOI-2762.01 & $<0.78$ & $-0.673_{-0.047}^{+0.055}$ &
\text{\sffamily X} &  
\text{\sffamily X} & 
\text{\sffamily X} & 
\text{\sffamily X} & 
- & ND & $<5.16$ \\ 
KOI-5284.01 & $<0.84$ & $+0.447_{-0.075}^{+0.073}$ &
\text{\sffamily X} & 
\checkmark & 
\text{\sffamily X} &   
N/A & N/A & ND & $<2.51$ \\ 
\hline
KOI-0072.02 & $<0.0025$ & $+0.313_{-0.016}^{+0.016}$ &
\checkmark & 
\checkmark & 
\text{\sffamily X} & 
N/A & N/A & ND & $<0.042^{\square}$ \\ 
- & $0.123_{-0.043}^{+0.043}$ & $+0.423_{-0.031}^{+0.036}$ &
\checkmark & 
\checkmark & 
\checkmark & 
N/A & N/A & FP-P$^{\diamond}$ & - \\ 
KOI-0245.01 & $<0.033$ & $+0.329_{-0.017}^{+0.016}$ &
\checkmark & 
\checkmark & 
\text{\sffamily X} &  
\text{\sffamily X} &
- & ND & $<0.18$ \\ 
KOI-0386.02 & $<0.0059$ & $+0.434_{-0.030}^{+0.027}$ &
\checkmark &
\checkmark & 
\text{\sffamily X} &
\text{\sffamily X} & 
- & ND & $<0.046$ \\ 
KOI-0518.02 & $<0.011$ & $-0.370_{-0.035}^{+0.042}$ &
\text{\sffamily X} & 
\text{\sffamily X} & 
\text{\sffamily X} & 
\text{\sffamily X} & 
- & ND & $<0.046$ \\ 
- & $<0.92$ & $+0.80_{-0.24}^{+0.13}$ &
\text{\sffamily X} & 
\checkmark & 
\text{\sffamily X} & 
\checkmark & 
\text{\sffamily X} & 
ND & $<4.02$ \\ 
KOI-0722.01 & $0.057_{-0.020}^{+0.025}$ & $+0.584_{-0.055}^{+0.047}$ &
\text{\sffamily X} & 
\checkmark & 
\checkmark &   
\text{\sffamily X} & 
- & FP-U & - \\ 
KOI-1783.01 & $<0.030$ & $-0.169_{-0.029}^{+0.034}$ &
\checkmark &
\text{\sffamily X} &  
\text{\sffamily X} & 
N/A & N/A & ND & $<1.21$ \\ 
KOI-2358.01 & $<0.015$ & $+0.543_{-0.054}^{+0.056}$ &
\checkmark & 
\checkmark & 
\text{\sffamily X} & 
\text{\sffamily X} & 
- & ND & $<0.090$ \\ 
KOI-3663.01 & $0.0079_{-0.0030}^{+0.0035}$ & $+0.121_{-0.011}^{+0.013}$ &
\checkmark & 
\checkmark & 
\checkmark & 
N/A & N/A & FP-I & - \\ 
KOI-3681.01 & $<0.0029$ & $+0.121_{-0.231}^{+0.006}$ &
\checkmark &  
\checkmark & 
\text{\sffamily X} &  
N/A & N/A & ND & $<1.84$ \\ [1ex]
\hline 
\multicolumn{10}{l}{N/A: Test not available, since all observed transits were used in basic analysis.} \\
\multicolumn{10}{l}{$^{\ast}$: The 95.45\% quantile is quoted for $(M_S/M_P)$ if B4a is not passed.} \\
\multicolumn{10}{l}{$^{\dagger}$: Strong evidence for TTVs; null fit was $\mathcal{H}_{S,R0}$ rather 
than $\mathcal{H}_{P}$} \\
\multicolumn{10}{l}{$^{\diamond}$: Classified as a false-positive since $\rho_P$ from moon model is 
incompatible with independently measured $\rho_P$} \\
\multicolumn{10}{l}{$^{\square}$: Empirical planetary mass available from radial velocities}
\end{tabular}
\label{tab:vetting} 
\end{table*}

\begin{figure*}
\begin{center}
\includegraphics[width=19.0 cm,angle=0,clip=true]{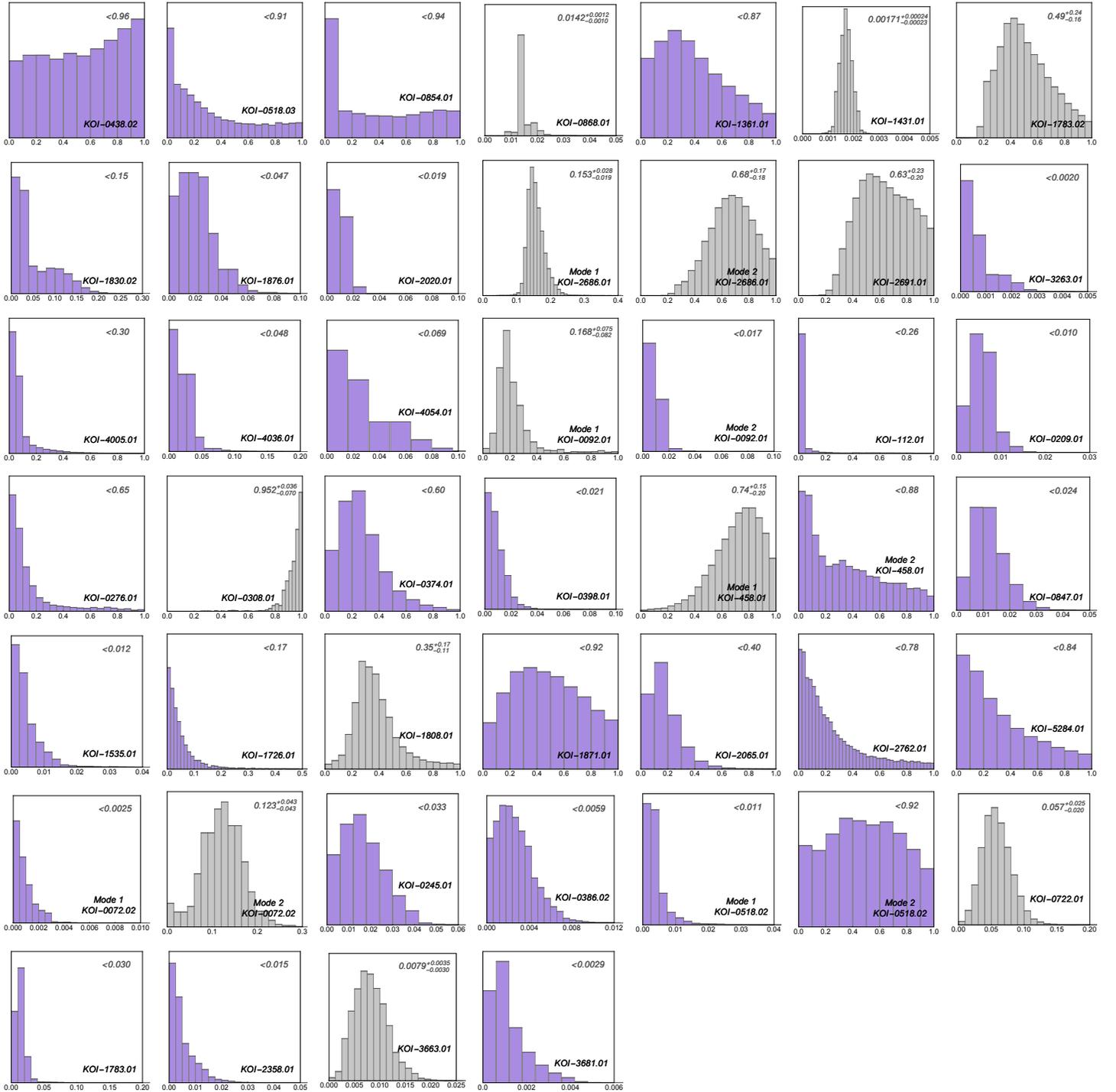}
\caption{
Histograms of the parameter $(M_S/M_P)$ for the 41 KOIs surveyed in this work.
Gray histograms are false-positive exomoon signals driven by some other 
perturber, colored ones are null detections. The estimated value of $(M_S/M_P)$
is provided in the upper-right corner of each panel.
} 
\label{fig:Msp_histos}
\end{center}
\end{figure*}

In Table~\ref{tab:vetting}, we summarize the results of our survey for each
KOI, providing the 68.3\% credible intervals for $(M_S/M_P)$ and $(R_S/R_P)$
computed from our marginalized posteriors. We also detail which detection
criteria were passed/failed and assign a classification (see 
\S\ref{sub:classification}) to each object. For some KOIs, two rows are shown
due to two distinct modes in the parameter $(M_S/M_P)$. In 
Figure~\ref{fig:Msp_histos}, we show the marginalized posterior of $(M_S/M_P)$ 
for each KOI (again showing all modes if more than one present).

\subsubsection{Kepler-10c}

For KOI-0072.02, also known as Kepler-10c, we elected to use all of the \kepler\ 
data, rather than the usual 75\% strategy discussed in \S\ref{sub:vetting}. This 
is because KOI-0072.02 is rare in providing us access to an additional vetting 
test. Specifically, a radial velocity solution exists for this object 
\citep{dumusque:2014}, allowing us to compare the radial velocity derived 
planetary density, $\rho_P$, to that from our photodynamical fits. We find two 
modes for KOI-0072.02 for which the first is a clear null detection, failing 
criterion B4a. The second mode passes all of our basic detection criteria 
though, and here the $\rho_P$ test is of great value. The second mode requires 
$\rho_P = 17.1_{-1.3}^{+2.0}$\,g\,cm$^{-3}$, which is highly incompatible with
the radial velocity measurement of \citet{dumusque:2014} of 
$\rho_P = 7.1\pm1.0$\,g\,cm$^{-3}$. This mode is therefore identified as a 
false-positive. Furthermore, as discussed in \S\ref{sub:TTVs}, KOI-0072.01
shows evidence for TTVs at the 5.0\,$\sigma$ level, meaning that the likely 
origin of this spurious detection is perturbations (i.e. FP-P). The 
consequences of TTVs in the Kepler-10 system are discussed in \S\ref{sub:TTVs}.

\subsubsection{PH-2b}

\begin{figure*}
\begin{center}
\includegraphics[width=19.0 cm,angle=0,clip=true]{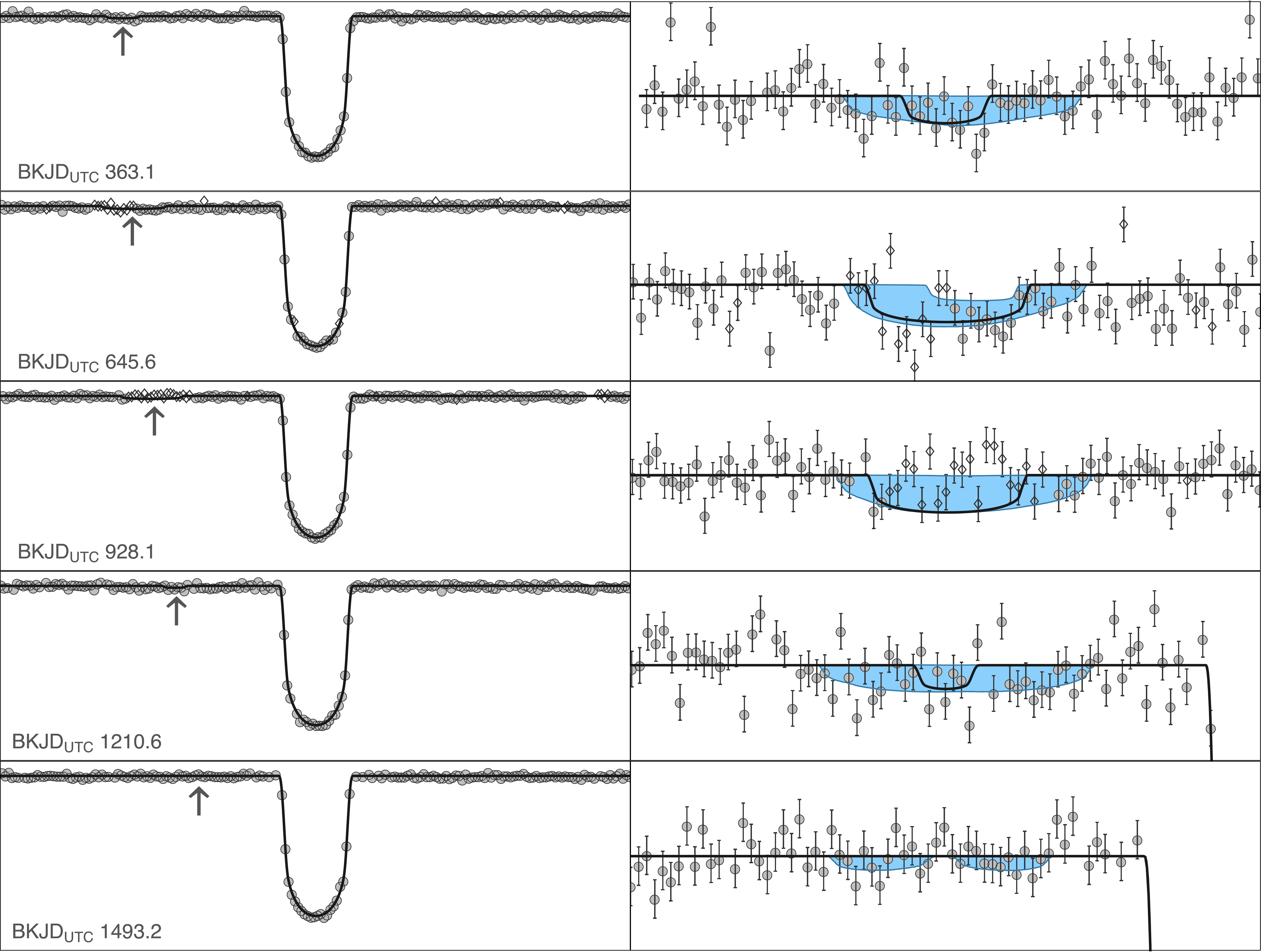}
\caption{
Transit light curves of all five epochs (each row) of PH-2b along with the 
maximum a-posteriori realization from model $\mathcal{H}_S$ (black line). 
Right panels show a zoom-in within $\pm0.8$\,d of the marked point of the left 
panel. Open diamond data points have zero weight in our fit as they have 
non-zero SAP QUALITY flags. The blue region denotes the 95.45\% confidence 
interval of our model. The low significance of the model, plus coincidental 
location of two of the moon signals with instrumental effects, leads to us 
reject this candidate.
} 
\label{fig:PH2b}
\end{center}
\end{figure*}

For KOI-3663.01, also known as PH-2b, all of the basic detection criteria are
passed but no follow-up tests are possible due to the small number of transit
epochs available (just five). Plotting the maximum a-posteriori realization from
model $\mathcal{H}_S$ reveals a solution driven by weak evidence for moon-like
transits in the first four transit epochs (see Figure~\ref{fig:PH2b}). As 
discussed later in \S\ref{sub:TTVs}, TTVs are not favored for this planet, since 
model $\mathcal{H}_P$ is preferred over $\mathcal{H}_{P,TTV}$ at the 5\,$\sigma$ 
level. As a result, the reality of the moon candidate rests solely on the 
reality of these moon-like transits.

The third moon-like transit, present in Q10, occurs almost perfectly in sync 
with a data gap caused by points being ignored by \cofiam, since the SAP QUALITY 
flags are non-zero (recall that we do not include such points in our analysis, 
see \S\ref{sub:data}). Specifically, from BKJD$_{\mathrm{UTC}}$\,926.93 to
BKJD$_{\mathrm{UTC}}$\,927.34, there is a reaction wheel zero crossing.
Without these points, there is in fact no data supporting the exomoon transit
predicted by model $\mathcal{H}_S$. Including the affected data show they are 
inconsistent with the model, as visible in Figure~\ref{fig:PH2b}.

For the second moon-like transit there is also a data gap obscuring the expected 
position of the moon's ingress signal. Remarkably, as with the aforementioned 
case, these points were also ignored by \cofiam\ due to another reaction wheel
zero crossing event. Including the affected points shows that these data are 
consistent with model $\mathcal{H}_S$. However, it should be cautioned that
low amplitude non-periodic transit signals, such as this, are known to be
reproducible by instrumental artifacts (e.g. see \citealt{KEP90}).

A final blow to the exomoon hypothesis comes from different detrendings of the
same time series. For example, repeating \cofiam\ on the PDCSAP data
and comparing the maximum a-posteriori model from $\mathcal{H}_S$ to the PDCSAP
detrended data reveals negligible improvement in the $\chi^2$ versus that of 
model $\mathcal{H}_P$. Specifically, we find $(\chi_{P}^2 - \chi_{S}^2) = +1.7$ 
for 3560 data points used in our analysis. We repeated this test using a variant 
of the Trend Filtering Algorithm (TFA) designed for \kepler\ and described in 
\citet{huang:2013}. Here, the results are even worse, with the planet model
$\mathcal{H}_P$ giving a superior explanation of the data with
$(\chi_{P}^2 - \chi_{S}^2) = -83.4$. As a result, we consider this candidate
to be an instrumental false-positive, reminiscent of the case of Kepler-90g.01
discussed in \citet{KEP90}.

\subsection{Transit Timing Variations}
\label{sub:TTVs}

\subsubsection{Method}

\begin{figure*}
\begin{center}
\includegraphics[width=18.0 cm,angle=0,clip=true]{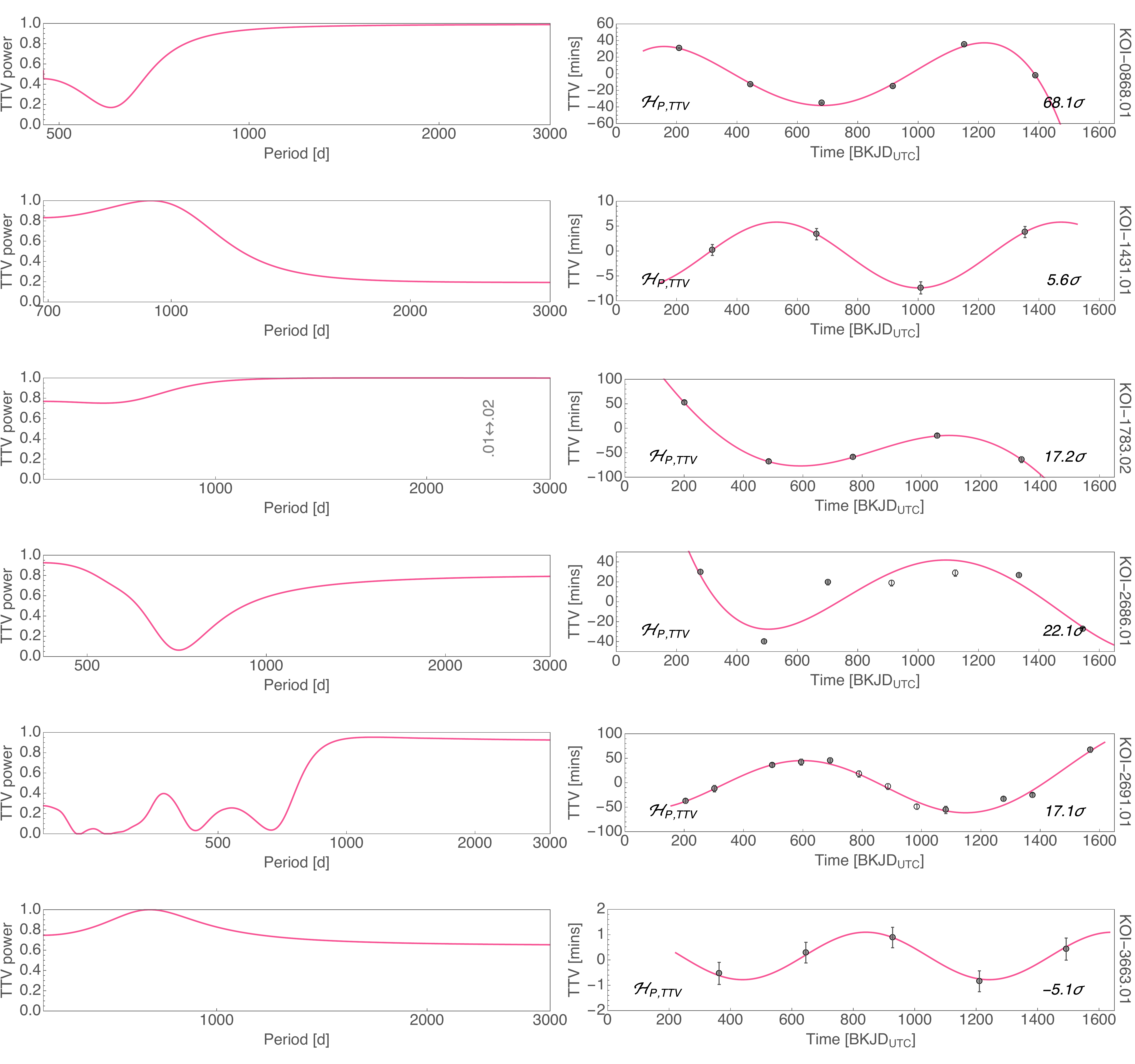}
\caption{
Right column shows the TTVs computed for 6 of the 12 KOIs which pass criteria 
B4a. These 6 KOIs have $N<17$ epochs enabling us to conduct the global fit, 
$\mathcal{H}_{P,TTV}$. The left panel shows the corresponding Lomb-Scargle 
periodogram with a BIC-based definition of power. The favored trial is plotted 
over the TTV data on the right panel, with the significance denoted in the 
corner. Negative significances indicate that the TTV model is disfavored.
} 
\label{fig:TTV_T}
\end{center}
\end{figure*}

We find that 12 of the 41 KOIs surveyed pass criterion B4a, suggestive of a
perturbation acting on these objects. To evaluate whether significant TTVs are 
present for these KOIs, we fit an additional planet model with freely varied
transit times, $\mathcal{H}_{P,TTV}$. This model has the same 5 basic shape
parameters for usual planets fits (see Table~\ref{tab:priors}) with the 
global epoch term ($\tau$) and period ($P$) now fixed to the maximum 
a-posteriori result from model $\mathcal{H}_{P}$. Additionally, for each of the
$N$ transits, an extra free parameter is included for that epoch's time of 
transit minimum, $\tau_i$, leading to $(5+N)$ free parameters. For each KOI, we 
evaluate the significance of the fit using the usual Bayesian model selection
method. The TTVs and the significance of the fit are shown in the right-hand
panels of Figure~\ref{fig:TTV_T}. We also compute a periodogram using the method
described in \citet{HEK3} (essentially a Lomb-Scargle periodogram with a 
BIC-based definition of power) and plot the highest power signal over the TTV 
data.

\begin{figure*}
\begin{center}
\includegraphics[width=18.0 cm,angle=0,clip=true]{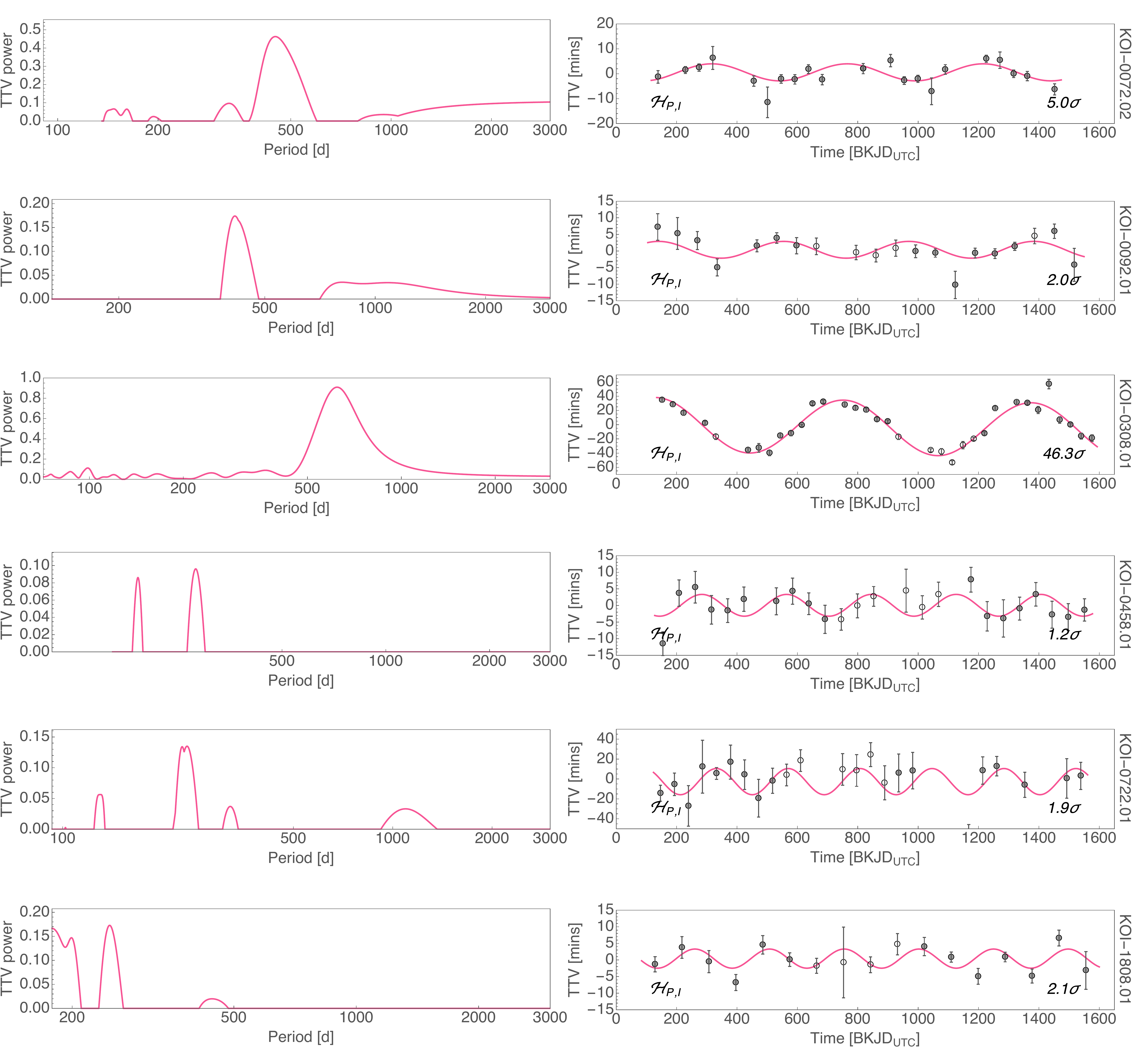}
\caption{
Right column shows the TTVs computed for 6 of the 12 KOIs which pass criteria 
B4a. These 6 KOIs have $N\geq17$ epochs and thus have too many free parameters 
for the global fit, $\mathcal{H}_{P,TTV}$, leading us to use a model of $N$ 
independent fits, $\mathcal{H}_{P,I}$. The left panel shows the corresponding 
Lomb-Scargle periodogram with a BIC-based definition of power. The favored trial 
is plotted over the TTV data on the right panel, with the significance denoted
in the corner.
} 
\label{fig:TTV_I}
\end{center}
\end{figure*}

Since the speed of \multi\ scales poorly with dimensionality 
\citep{handley:2015}, we find that KOIs with $N\geq17$ were not practical to fit 
using model $\mathcal{H}_{P,TTV}$. Instead, we fit each transit independently 
fixing the limb darkening coefficients to theoretical values interpolated from 
the \citet{claret:2011} tabulations and the period, $P$, as before, leaving 4 
free parameters per fit. We refer to this model as $\mathcal{H}_{P,I}$ and show 
the results in Figure~\ref{fig:TTV_I}. Since the summed Bayesian evidence now 
accounts for excessive freedom, such as duration and depth changes, we instead 
define a significance for the TTV fit using the Bayesian Information Criterion 
between the maximum periodogram peak signal and a null fit. These significances 
are not as statistically robust as the Bayesian evidence method, but serve as a 
useful proxy.

\subsubsection{Dynamically Active KOIs}

Seven of twelve systems inspected for TTVs show $\geq5$\,$\sigma$ evidence for
TTVs. Five of these show extremely strong evidence ($>17$\,$\sigma$) for which 
there can be little doubt the KOIs are being perturbed; specifically
KOI-0308.01, KOI-0868.01, KOI-1783.02, KOI-2686.01 and KOI-2691.01. Of these
five, only KOI-1783.02 is known to reside in a multiple transiting planet
system and the TTV period is compatible with a perturbation from KOI-1783.01,
although there is insufficient data to claim so unambiguously. The other four
strongly perturbed KOIs are in single KOI systems, indicating at least four
additional massive bodies not known to transit the parent star.

Of the two marginal cases for TTVs, KOI-0072.02 and KOI-1431.01, the former is
of particular interest being that this is Kepler-10c, the claimed solid
Neptune-mass planet measured through radial velocities by \citet{dumusque:2014}.
Kepler-10b is essentially decoupled from 10c and is not expected to perturb 10c, 
suggesting an additional unknown massive body in the system. This object is 
likely to be of planetary mass to have avoided detection by radial velocities 
thus far. The period and mass of this object cannot be uniquely inferred from a 
single sinusoidal-like set of TTVs \citep{nesvorny:2014,deck:2014}. However, a 
near mean motion resonance configuration is most probable due to the fact that
this configuration leads to the largest amplitude effects. Further, a planet on
an exterior orbit would have a greater chance of having not been seen transiting
to date, based on the system geometry.

Whilst the evidence for TTVs in Kepler-10c is not decisive, it is worth noting 
that the radial velocity periodogram (see \citealt{dumusque:2014}) shows 
numerous additional peaks to the frequencies of
10b and 10c. Notably, the next largest peak above the 0.1\% false alarm 
probability threshold occurs at 158.54\,d, almost coincidental with a 7:2 ratio
with 10c and potentially responsible for a near mean motion resonance 
interaction. We recommend additional radial velocities to identify the cause
of the observed TTVs.

\subsection{Refined Planet Parameters}
\label{sub:refined}

\begin{figure}
\begin{center}
\includegraphics[width=8.4 cm,angle=0,clip=true]{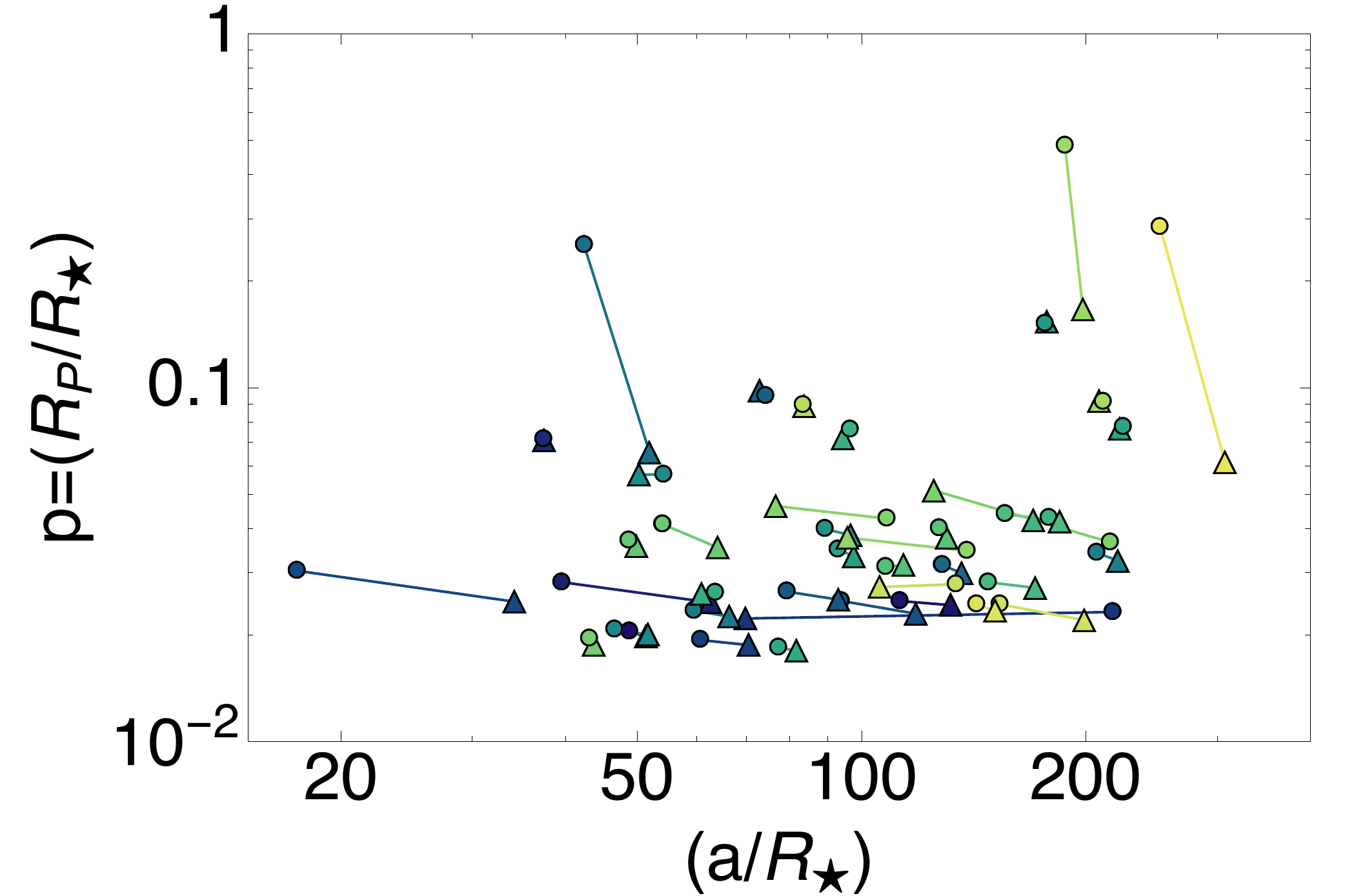}
\caption{
Comparison of the NASA Exoplanet Archive listed transit parameters (triangles)
versus the values derived in this work (circles). Specific KOIs are connected
by a line and have the same color coding. The plotted points represent the 
medians of the parameter posteriors.
} 
\label{fig:parameter_revisions}
\end{center}
\end{figure}

An additional useful result from this work are parameter posteriors derived for
each of the planets surveyed. Since no exomoons are found, we use model
$\mathcal{H}_P$ to estimate refined planetary parameters of each KOI studied.
In the cases of the seven dynamically active KOIs, we use the TTV based models
instead. The refined parameters are provided in Table~\ref{tab:planetparams}.
Figure~\ref{fig:parameter_revisions} illustrates the differences between the
parameters derived in this work and those on the NASA Exoplanet Archive, which
agree well in general.

\begin{table*}
\caption{\emph{Planet properties as derived using model $\mathcal{H}_P$.
}} 
\centering 
\begin{tabular}{c c c c c c c c} 
\hline
KOI & $p$ & $\log_{10}(\rho_{\star}$\,[kg\,m$^{-3}$]) & $b$ & $P$\,[d] & $\tau$\,[BJKD] & $q_1$ & $q_2$ \\ [0.5ex] 
\hline
KOI-0438.02 & $0.03147_{-0.00089}^{+0.00128}$ & $4.158_{-0.295}^{+0.086}$ & $0.36_{-0.25}^{+0.31}$ & $52.661599_{-0.000097}^{+0.000097}$ & $868.12003_{-0.00081}^{+0.00077}$ & $0.30_{-0.18}^{+0.34}$ & $0.25_{-0.18}^{+0.36}$ \\
KOI-0518.03 & $0.03415_{-0.00088}^{+0.00229}$ & $3.44_{-0.39}^{+0.10}$ & $0.40_{-0.28}^{+0.33}$ & $247.3545_{-0.0013}^{+0.0012}$ & $858.8682_{-0.0017}^{+0.0017}$ & $0.48_{-0.23}^{+0.31}$ & $0.23_{-0.15}^{+0.31}$ \\
KOI-0854.01 & $0.0398_{-0.0016}^{+0.0020}$ & $3.63_{-0.40}^{+0.14}$ & $0.46_{-0.32}^{+0.31}$ & $56.05608_{-0.00020}^{+0.00021}$ & $873.8370_{-0.0018}^{+0.0020}$ & $0.27_{-0.19}^{+0.37}$ & $0.31_{-0.22}^{+0.38}$ \\
KOI-0868.01$^{\dagger}$ & $0.1516_{-0.0014}^{+0.0024}$ & $3.270_{-0.014}^{+0.019}$ & $0.816_{-0.019}^{+0.012}$ & $235.99726_{-0.00013}^{+0.00013}$ & $680.40440_{-0.00023}^{+0.00023}$ & $0.35_{-0.17}^{+0.30}$ & $0.28_{-0.20}^{+0.36}$ \\
KOI-1361.01 & $0.03488_{-0.00088}^{+0.00186}$ & $3.626_{-0.284}^{+0.085}$ & $0.37_{-0.25}^{+0.30}$ & $59.87792_{-0.00013}^{+0.00012}$ & $869.7208_{-0.0011}^{+0.0011}$ & $0.52_{-0.23}^{+0.29}$ & $0.28_{-0.17}^{+0.30}$ \\
KOI-1431.01$^{\dagger}$ & $0.0774_{-0.0014}^{+0.0026}$ & $3.254_{-0.032}^{+0.033}$ & $0.863_{-0.011}^{+0.010}$ & $345.15902_{-0.00035}^{+0.00035}$ & $633.46686_{-0.00043}^{+0.00042}$ & $0.52_{-0.12}^{+0.17}$ & $0.29_{-0.21}^{+0.38}$ \\
KOI-1783.02$^{\dagger}$ & $0.0428_{-0.0040}^{+0.0019}$ & $3.12_{-0.17}^{+0.52}$ & $0.844_{-0.285}^{+0.044}$ & $294.0729_{-0.0014}^{+0.0015}$ & $768.9686_{-0.0018}^{+0.0018}$ & $0.39_{-0.20}^{+0.36}$ & $0.31_{-0.21}^{+0.40}$ \\
KOI-1830.02 & $0.0440_{-0.0019}^{+0.0031}$ & $3.26_{-0.27}^{+0.18}$ & $0.52_{-0.33}^{+0.21}$ & $198.71078_{-0.00060}^{+0.00058}$ & $752.6515_{-0.0015}^{+0.0016}$ & $0.42_{-0.14}^{+0.25}$ & $0.55_{-0.27}^{+0.29}$ \\
KOI-1876.01 & $0.0401_{-0.0013}^{+0.0021}$ & $3.757_{-0.239}^{+0.086}$ & $0.36_{-0.25}^{+0.27}$ & $82.53428_{-0.00027}^{+0.00027}$ & $869.4001_{-0.0013}^{+0.0013}$ & $0.40_{-0.17}^{+0.29}$ & $0.60_{-0.28}^{+0.27}$ \\
KOI-2020.01 & $0.0370_{-0.0034}^{+0.0039}$ & $2.25_{-0.40}^{+0.39}$ & $0.71_{-0.43}^{+0.15}$ & $110.9650_{-0.0010}^{+0.0011}$ & $852.8362_{-0.0043}^{+0.0045}$ & $0.62_{-0.24}^{+0.25}$ & $0.43_{-0.26}^{+0.31}$ \\
KOI-2686.01$^{\dagger}$ & $0.04053_{-0.00066}^{+0.00113}$ & $3.804_{-0.145}^{+0.054}$ & $0.30_{-0.20}^{+0.22}$ & $211.03338_{-0.00055}^{+0.00051}$ & $912.2156_{-0.0012}^{+0.0012}$ & $0.39_{-0.15}^{+0.25}$ & $0.44_{-0.21}^{+0.28}$ \\
KOI-2691.01$^{\dagger}$ & $0.04262_{-0.00091}^{+0.00156}$ & $3.400_{-0.196}^{+0.070}$ & $0.33_{-0.22}^{+0.26}$ & $97.44645_{-0.00028}^{+0.00028}$ & $886.9525_{-0.0012}^{+0.0012}$ & $0.28_{-0.11}^{+0.21}$ & $0.65_{-0.30}^{+0.24}$ \\
KOI-3263.01 & $0.48_{-0.29}^{+0.37}$ & $4.324_{-0.055}^{+0.064}$ & $1.28_{-0.37}^{+0.39}$ & $76.879333_{-0.000046}^{+0.000046}$ & $761.91724_{-0.00025}^{+0.00025}$ & $0.27_{-0.20}^{+0.37}$ & $0.46_{-0.32}^{+0.36}$ \\
KOI-4005.01 & $0.02757_{-0.00093}^{+0.00098}$ & $3.14_{-0.29}^{+0.10}$ & $0.39_{-0.26}^{+0.29}$ & $178.1388_{-0.0013}^{+0.0015}$ & $744.5337_{-0.0035}^{+0.00035}$ & $0.14_{-0.10}^{+0.24}$ & $0.40_{-0.29}^{+0.38}$ \\
KOI-4036.01 & $0.0244_{-0.0023}^{+0.1016}$ & $3.38_{-1.50}^{+0.29}$ & $0.62_{-0.42}^{+0.46}$ & $168.8102_{-0.0013}^{+0.0015}$ & $886.6796_{-0.0041}^{+0.0040}$ & $0.66_{-0.31}^{+0.23}$ & $0.59_{-0.34}^{+0.27}$ \\
KOI-4054.01 & $0.02439_{-0.00084}^{+0.00094}$ & $3.28_{-0.32}^{+0.10}$ & $0.39_{-0.27}^{+0.31}$ & $169.1330_{-0.0011}^{+0.0013}$ & $878.3472_{-0.0035}^{+0.0029}$ & $0.19_{-0.13}^{+0.29}$ & $0.33_{-0.24}^{+0.40}$ \\
\hline
KOI-0092.01 & $0.0249_{-0.0018}^{+0.0017}$ & $3.80_{-0.30}^{+0.38}$ & $0.71_{-0.42}^{+0.12}$ & $65.704565_{-0.000054}^{+0.000052}$ & $794.48803_{-0.00032}^{+0.00032}$ & $0.52_{-0.16}^{+0.32}$ & $0.17_{-0.11}^{+0.31}$ \\
KOI-0112.01 & $0.0282_{-0.0015}^{+0.0010}$ & $2.65_{-0.13}^{+0.29}$ & $0.784_{-0.164}^{+0.045}$ & $51.079310_{-0.000061}^{+0.000061}$ & $900.29048_{-0.00057}^{+0.00057}$ & $0.338_{-0.071}^{+0.125}$ & $0.38_{-0.27}^{+0.38}$ \\
KOI-0209.01 & $0.07158_{-0.00035}^{+0.00040}$ & $2.584_{-0.026}^{+0.024}$ & $0.233_{-0.106}^{+0.073}$ & $50.790364_{-0.000021}^{+0.000021}$ & $846.69768_{-0.00019}^{+0.00019}$ & $0.232_{-0.034}^{+0.036}$ & $0.458_{-0.066}^{+0.081}$ \\
KOI-0276.01 & $0.01933_{-0.00050}^{+0.00089}$ & $3.39_{-0.27}^{+0.16}$ & $0.50_{-0.32}^{+0.21}$ & $41.745992_{-0.000033}^{+0.000034}$ & $961.82689_{-0.00037}^{+0.00036}$ & $0.44_{-0.12}^{+0.19}$ & $0.16_{-0.11}^{+0.19}$ \\
KOI-0308.01$^{\ast}$ & $0.02365_{-0.00020}^{+0.00020}$ & $2.477_{-0.036}^{+0.036}$ & $0.892_{-0.017}^{+0.017}$ & $35.597254_{-0.000044}^{+0.000043}$ & $899.47632_{-0.00056}^{+0.00057}$ & - & - \\
KOI-0374.01 & $0.0249_{-0.0011}^{+0.0013}$ & $2.72_{-0.27}^{+0.25}$ & $0.61_{-0.36}^{+0.16}$ & $172.70406_{-0.00051}^{+0.00049}$ & $755.0368_{-0.0013}^{+0.0013}$ & $0.45_{-0.12}^{+0.20}$ & $0.24_{-0.15}^{+0.26}$ \\
KOI-0398.01 & $0.0945_{-0.0018}^{+0.0020}$ & $3.461_{-0.045}^{+0.049}$ & $0.616_{-0.050}^{+0.039}$ & $51.846858_{-0.000028}^{+0.000028}$ & $844.09103_{-0.00022}^{+0.00022}$ & $0.75_{-0.21}^{+0.17}$ & $0.330_{-0.093}^{+0.166}$ \\
KOI-0458.01 & $0.25_{-0.18}^{+0.52}$ & $2.70_{-0.12}^{+0.23}$ & $1.18_{-0.24}^{+0.54}$ & $53.718027_{-0.000068}^{+0.000067}$ & $906.41442_{-0.00062}^{+0.00062}$ & $0.25_{-0.19}^{+0.37}$ & $0.57_{-0.38}^{+0.33}$ \\
KOI-0847.01 & $0.0568_{-0.0014}^{+0.0023}$ & $2.664_{-0.155}^{+0.099}$ & $0.41_{-0.27}^{+0.19}$ & $80.87199_{-0.00015}^{+0.00015}$ & $850.87236_{-0.00088}^{+0.00087}$ & $0.41_{-0.14}^{+0.21}$ & $0.42_{-0.17}^{+0.27}$ \\
KOI-1535.01 & $0.01844_{-0.00055}^{+0.00105}$ & $3.24_{-0.58}^{+0.10}$ & $0.40_{-0.28}^{+0.42}$ & $70.69782_{-0.00025}^{+0.00026}$ & $824.8510_{-0.0019}^{+0.0016}$ & $0.22_{-0.15}^{+0.30}$ & $0.20_{-0.16}^{+0.39}$ \\
KOI-1726.01 & $0.02632_{-0.00082}^{+0.00213}$ & $3.38_{-0.35}^{+0.13}$ & $0.45_{-0.30}^{+0.28}$ & $44.964066_{-0.000055}^{+0.000056}$ & $819.02111_{-0.00060}^{+0.00061}$ & $0.44_{-0.17}^{+0.26}$ & $0.49_{-0.24}^{+0.31}$ \\
KOI-1808.01 & $0.02816_{-0.00095}^{+0.00296}$ & $3.88_{-0.63}^{+0.18}$ & $0.52_{-0.36}^{+0.34}$ & $89.193549_{-0.000080}^{+0.000078}$ & $931.37212_{-0.00048}^{+0.00047}$ & $0.39_{-0.16}^{+0.33}$ & $0.21_{-0.14}^{+0.36}$ \\
KOI-1871.01 & $0.0312_{-0.0012}^{+0.0026}$ & $3.440_{-0.264}^{+0.086}$ & $0.36_{-0.25}^{+0.29}$ & $92.72967_{-0.00033}^{+0.00035}$ & $826.5496_{-0.0020}^{+0.0018}$ & $0.57_{-0.20}^{+0.25}$ & $0.68_{-0.23}^{+0.21}$ \\
KOI-2065.01 & $0.041_{-0.013}^{+0.020}$ & $2.67_{-0.32}^{+1.26}$ & $0.942_{-0.620}^{+0.045}$ & $80.23201_{-0.00049}^{+0.00053}$ & $1044.9485_{-0.0026}^{+0.0027}$ & $0.73_{-0.34}^{+0.20}$ & $0.65_{-0.39}^{+0.25}$ \\
KOI-2762.01 & $0.0345_{-0.0014}^{+0.0021}$ & $3.45_{-0.44}^{+0.12}$ & $0.41_{-0.29}^{+0.36}$ & $132.99679_{-0.00090}^{+0.00099}$ & $923.3727_{-0.0031}^{+0.0030}$ & $0.33_{-0.23}^{+0.37}$ & $0.29_{-0.21}^{+0.37}$ \\
KOI-5284.01 & $0.28_{-0.23}^{+0.49}$ & $3.30_{-0.15}^{+1.12}$ & $1.21_{-0.68}^{+0.50}$ & $389.3128_{-0.0022}^{+0.0023}$ & $1129.9934_{-0.0019}^{+0.0018}$ & $0.56_{-0.39}^{+0.30}$ & $0.63_{-0.43}^{+0.27}$ \\
\hline
KOI-0072.02$^{\ast}$ & $0.020414_{-0.000062}^{+0.000062}$ & $3.0286_{-0.0016}^{+0.0016}$ & $0.3182_{-0.0063}^{+0.0063}$ & $45.294314_{-0.000031}^{+0.000032}$ & $908.68114_{-0.00027}^{+0.00028}$ & - & - \\
KOI-0245.01 & $0.023201_{-0.000059}^{+0.000062}$ & $5.088_{-0.012}^{+0.012}$ & $0.5350_{-0.0086}^{+0.0084}$ & $39.792238_{-0.000011}^{+0.000011}$ & $772.133411_{-0.000099}^{+0.000100}$ & $0.382_{-0.028}^{+0.030}$ & $0.436_{-0.056}^{+0.057}$ \\
KOI-0386.02 & $0.0265_{-0.0011}^{+0.0026}$ & $3.21_{-0.59}^{+0.19}$ & $0.53_{-0.36}^{+0.32}$ & $76.73312_{-0.00034}^{+0.00035}$ & $891.2658_{-0.0019}^{+0.0019}$ & $0.38_{-0.21}^{+0.33}$ & $0.30_{-0.21}^{+0.39}$ \\
KOI-0518.02 & $0.02342_{-0.00080}^{+0.00135}$ & $3.32_{-0.44}^{+0.14}$ & $0.47_{-0.32}^{+0.31}$ & $44.00035_{-0.00013}^{+0.00012}$ & $870.7686_{-0.0015}^{+0.0014}$ & $0.27_{-0.16}^{+0.30}$ & $0.30_{-0.22}^{+0.40}$ \\
KOI-0722.01 & $0.02070_{-0.00056}^{+0.00105}$ & $2.95_{-0.45}^{+0.13}$ & $0.45_{-0.31}^{+0.33}$ & $46.40648_{-0.00015}^{+0.00015}$ & $192.9939_{-0.0022}^{+0.0021}$ & $0.21_{-0.12}^{+0.22}$ & $0.29_{-0.22}^{+0.40}$ \\
KOI-1783.01 & $0.0761_{-0.0035}^{+0.0045}$ & $2.972_{-0.051}^{+0.085}$ & $0.916_{-0.025}^{+0.015}$ & $134.47864_{-0.00013}^{+0.00013}$ & $841.67398_{-0.00040}^{+0.00038}$ & $0.62_{-0.35}^{+0.29}$ & $0.48_{-0.34}^{+0.37}$ \\
KOI-2358.01 & $0.01955_{-0.00078}^{+0.00241}$ & $2.68_{-0.64}^{+0.15}$ & $0.47_{-0.32}^{+0.38}$ & $56.49384_{-0.00034}^{+0.00034}$ & $991.3899_{-0.0033}^{+0.0037}$ & $0.54_{-0.27}^{+0.30}$ & $0.31_{-0.21}^{+0.34}$ \\
KOI-3663.01 & $0.09099_{-0.00051}^{+0.00046}$ & $3.347_{-0.017}^{+0.019}$ & $0.276_{-0.062}^{+0.045}$ & $282.525517_{-0.000094}^{+0.000093}$ & $928.12271_{-0.00013}^{+0.00013}$ & $0.426_{-0.043}^{+0.049}$ & $0.316_{-0.033}^{+0.037}$ \\
KOI-3681.01 & $0.08942_{-0.00013}^{+0.00013}$ & $2.3654_{-0.0056}^{+0.0058}$ & $0.215_{-0.023}^{+0.020}$ & $217.831780_{-0.000050}^{+0.000050}$ & $892.315527_{-0.000100}^{+0.000096}$ & $0.279_{-0.012}^{+0.012}$ & $0.381_{-0.016}^{+0.017}$ \\ [1ex]
\hline 
\multicolumn{8}{l}{$^{\dagger}$: $\mathcal{H}_{P,TTV}$ used instead due to significant TTVs present} \\
\multicolumn{8}{l}{$^{\ast}$: $\mathcal{H}_{P,I}$ used instead due to significant TTVs present
and $\geq17$ transit epochs} \\
\end{tabular}
\label{tab:planetparams} 
\end{table*}

\section{DISCUSSION}
\label{sec:discussion}

\subsection{Overview}
\label{sub:overview}

With the 41 KOIs surveyed in this work, the total number of unique KOIs surveyed
using Bayesian photodynamics is now 57. In every case, the analysis comes from
the HEK project and were reported in \citet{KOI872} and \citet{HEK2,HEK3,HEK4,
KEP90}. In this survey, as with the previous smaller samples, we find no 
compelling evidence for an exomoon.

Despite the sizable number of objects surveyed, we caution the community against 
attempting to derive $\eta_{\leftmoon}$ constraints (the occurrence rate of 
exomoons) at this time, since some of our most interesting candidates remain 
under detailed analysis and are yet to be presented. This is an inevitable
consequence of the fact that putative candidates require in-depth vetting,
such as the case with Kepler-90g.01 where pixel-level analysis was required
\citep{KEP90}.

\subsection{Unknown False Positives}
\label{sub:FPU}

Of the 41 KOIs studied, we find four instances of false-positives of unknown 
origin; KOI-0092.01, KOI-0458.01, KOI-0722.01 and KOI-1808.01. For the first 
three, criteria B2 and B4a are passed and for KOI-1808.01 all of the basic 
criteria are satisfied. However, all four KOIs fail the follow-up test, F2a. 
This means that a Keplerian moon model provides a worse prediction (in a 
$\chi^2$ sense) of the $\sim25$\% of ignored transits than a vanilla planet-only 
model, indicating that a moon cannot explain the observations.

When dealing with a handful of transits, quasi-periodic distortions to the 
transit profile, such as those due to spots \citep{beky:2014}, can be well 
fitted by the flexible exomoon model. However, since an exomoon is not the 
underlying cause, this model lacks any predictive power and thus should fail 
F2a. We therefore suggest that stellar activity is likely responsible for these 
four instances. As an example, we briefly discuss the most promising case of 
KOI-1808.01 which passes all of the basic criteria. Inspection of the posteriors 
reveals that the planet-moon separation is required to be tight at 
$8.2_{-2.2}^{+2.2}$ planetary radii leading to no clean, separated moon transits 
in the best fit. Instead, the signal is driven by distortions to the planetary 
transit event, consistent with the effects of spots.

\subsection{TTV Systems}
\label{sub:TTV}

Although it is not the objective of our work, we find seven systems classed
as FP-P, that is exomoon false-positives due to perturbations from another body.
These cases display strong evidence for TTVs ($\geq5$\,$\sigma$) and account
for $17$\% of the KOIs surveyed. Curiously, five of these instances occur
for the 16 targets selected through TSA, which has a coincidental probability
of occurring of just $7.7$\%. Since the TSA targets were selected to be in the
habitable zone, they are more biased towards long period KOIs, perhaps 
suggesting that TTVs are more common for such worlds. Whatever the cause, based 
on the 41 KOIs studied here, habitable zone TSA planetary candidates have an 
a-priori chance of displaying TTVs of $(31\pm12)$\,\%. 

This rate is substantially higher than that reported by \citet{mazeh:2013}, who 
find 130 KOIs with significant TTVs out of a total sample of 1897 i.e. 
$(6.9\pm0.6)$\,\%. This is incompatible with the rate seen for the 
TSAs beyond the inner edge of the habitable zone (as defined by 
\citealt{zsom:2013}) at a level of 3.2\,$\sigma$, meaning coincidental sample 
selection is highly improbable to be responsible. However, the sample studied by 
\citet{mazeh:2013} casts a much wider net than the carefully selected targets of 
TSA. Specifically, this rate is for the 2321 KOIs in the \citet{borucki:2011,
borucki:2012} catalogs with a SNR$>2.5$ (giving 1960 KOIs) exhibiting 7 or more 
transits in the first 12 quarters of data (giving 1897 KOIs). The latter filter 
excludes long period planets ($P\lesssim150$\,days) and moreover the sample is 
heavily biased towards short period events due the higher detection probability 
($\sim P^{-5/3}$; \citealt{gaudi:2006}).

The tension between these two rates could be resolved by a period dependency for 
the occurrence rate of detectable TTVs. Our TSA sample is designed to pick
gaseous planets in the habitable zone or further out, implying that habitable
zone gaseous planets are frequently accompanied by additional planets, likely in
period commensurabilities. These hidden planets probably reside in exterior 
orbits, in order to explain having not been seen to transit (since multi-planet 
systems are typically coplanar; \citealt{fabrycky:2014}). However, we caution 
that our TSA method uses a complex series of criteria, such as invoking tidal 
evolution models, which may make them not representative of habitable zone 
gaseous planets. The occurrence rate of dynamically active habitable zone or
cooler gaseous planets is therefore an interesting question to pursue in future
work.

\subsection{Empirical Sensitivity to Exomoons}
\label{sub:sensitivity}

Several previous works have attempted to estimate the expected sensitivity of 
\kepler\ and other telescopes to exomoons (e.g. \citealt{kippingetal:2009}; 
\citealt{simon:2009}). None of these works have gauged the sensitivity of full
photodynamics, rather focusing on just one or two of the multitude of effects by
which an exomoon may betray its presence. Further, real noise structure and 
complex inter-parameter degeneracies have not been realistically modeled. The
most robust estimate of our photometric sensitivity to exomoons comes from
examining the empirical upper limits derived from a reasonably large sample,
such as those from the HEK project surveys.

\begin{figure}
\includegraphics[width=8.6cm,angle=0,clip=true]{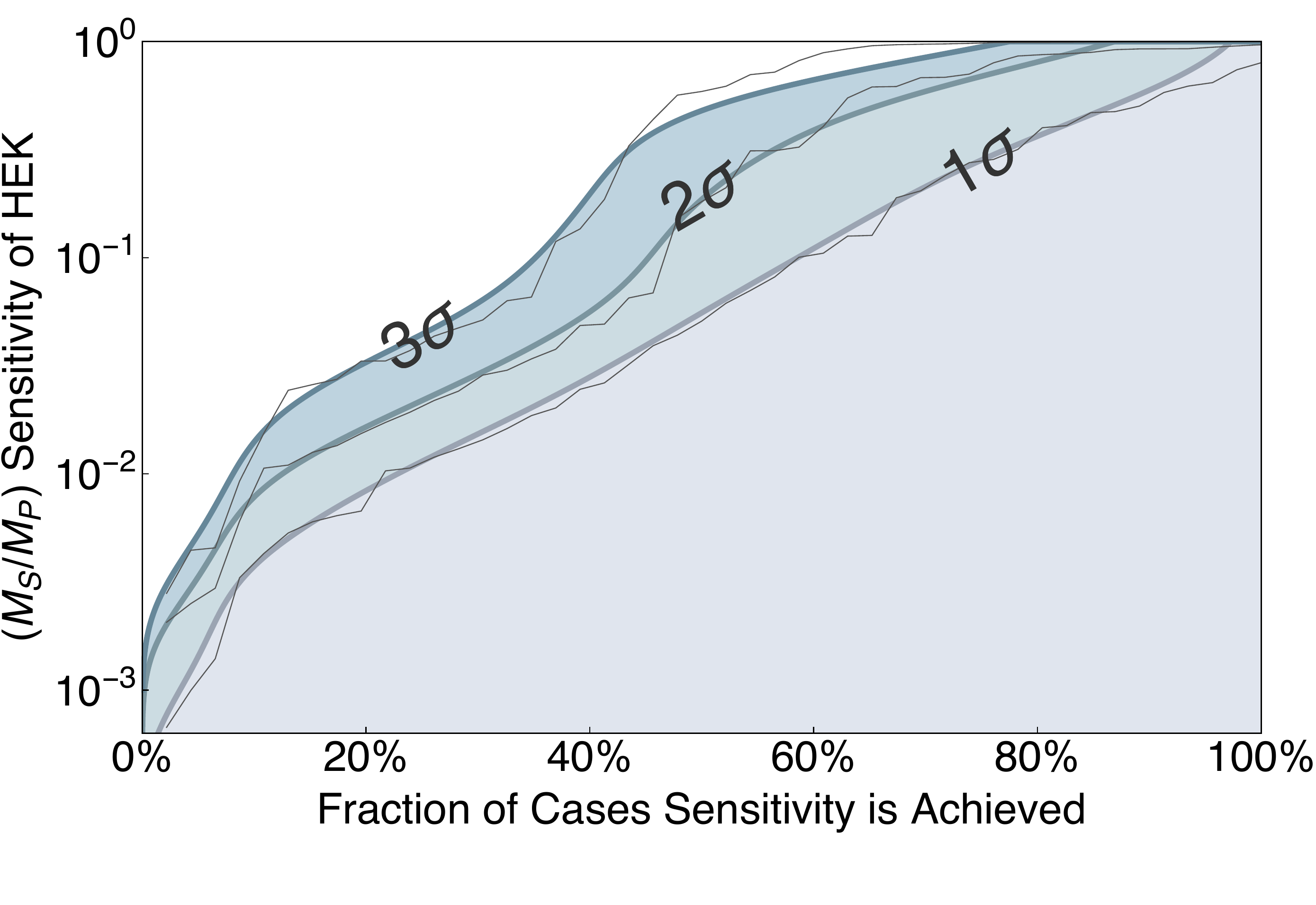}\\
\includegraphics[width=8.6cm,angle=0,clip=true]{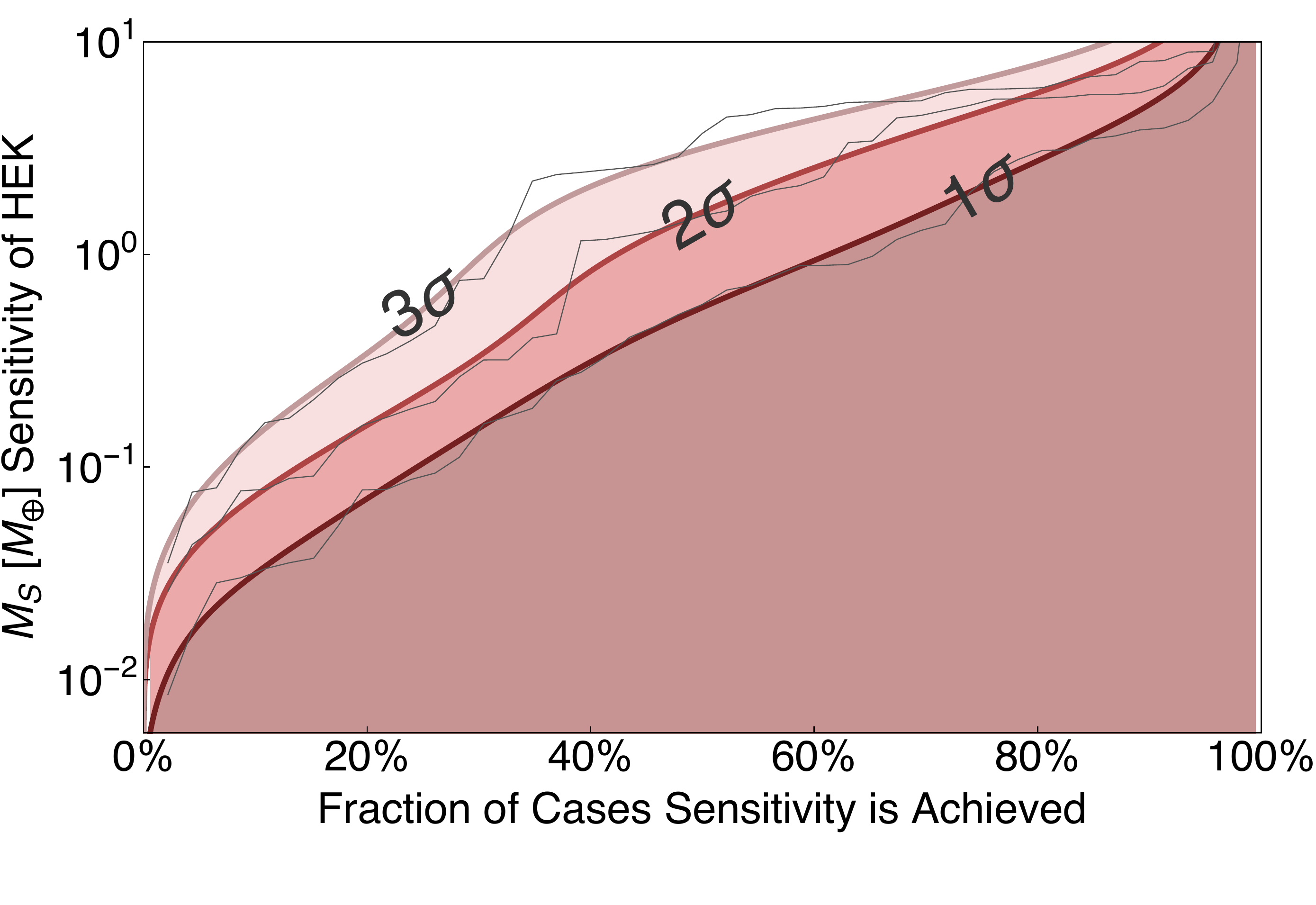}
\caption{
\textbf{Upper:} Smoothed empirical sensitivity of the HEK survey to the 
satellite-to-planet mass-ratio. Based on 46 null results (thin gray line) drawn 
from the 57 unique KOIs studied. HEK is sensitive to Earth-Moon mass-ratios 
for $\simeq15$\% of null cases and to Pluto-Charon mass-ratios for 
$\simeq47$\% (based on 95.45\% confidences).
\textbf{Lower:} Same as top, except sensitivity quoted in absolute masses via
estimated planetary masses. HEK is sensitive to 0.3\,$M_{\oplus}$ mini-Earths
for $\simeq30$\% of cases and to Earths for $\simeq40$\% (based on 95.45\% 
confidences).
}
\label{fig:sensitivity}
\end{figure}

Of the 57 unique KOIs studied, we find 46 null detections which are suitable for
deriving upper limits from. This indicates that we obtain null detections
for $(81\pm5)$\% of cases studied, with the others being false-positives. From
the null detections, we find a diverse range of sensitivities in $(M_S/M_P)$,
ranging from $1.9\times10^{-3}$ (95.45\% upper limit of KOI-2020.01) to nearly 
unity, as shown in Figure~\ref{fig:sensitivity} (for reference, the Earth-Moon
mass-ratio is $12.3\times10^{-3}$). This diversity is likely due to 
a complex combination of factors, such as the target magnitude, planetary 
transit depth and duration, correlated noise structure, Hill sphere size, 
latent TTVs from perturbing planets, photometric cadence and number of transits. 
Regardless, we are able to infer that, assuming our sample is typical, the null 
detections are sensitive to Earth-Moon mass-ratios or better for $\simeq15$\% of 
cases and to Pluto-Charon mass-ratios for $\simeq47$\% (based on 95.45\% 
confidences).

The relative mass limits of $(M_S/M_P)$ may be converted to absolute masses 
limits in $M_S$ by estimating $M_P$. For cases such as Kepler-10c, the physical
mass is known \citep{dumusque:2014}, and here our upper limit corresponds to
1.7 Ganymede masses (at 95.45\% confidence), demonstrating our ability to find
moons akin to those in the Solar System. In the other cases, we must assign a 
mass based on the observed radius, $R_P$. For self-consistency, we use the same
$R_P$-$M_P$ relation used for TSA, namely that discussed in \citet{HEK4}, which
is predominantly an empirical relationship. As shown in the right panel of
Figure~\ref{fig:sensitivity}, this allows to estimate that we are sensitive to
Earth-like moons or smaller for $\simeq40$\% of the null cases. The lowest mass 
exomoon that can be considered capable of supporting an Earth-like atmosphere is 
thought to be at 0.3\,$M_{\oplus}$ \citep{raymond:2007}, for which we find
sensitivity for $\simeq30$\% of null cases.

In summary then, based on empirical sensitivity limits, we show for the first
time that the HEK project is sensitive to even the smallest moons capable of
being Earth-like for 1 in 4 cases (after accounting for false-positives). In 
terms of planet-mass ratios, we find even that the Earth-Moon mass-ratio is 
detectable for 1 in 8 of cases, posing a challenge but not an insurmountable
barrier. Mass ratios of $\sim10^{-4}$, such as that of the Galilean satellites, 
have never been achieved. However, if Galilean-like satellites reside around 
lower-mass planets than Jupiter, of order $\sim20$\,$M_{\oplus}$, then we do
find sensitivity, as demonstrated by the limit of 1.7 Ganymede masses achieved 
for Kepler-10c.

\subsection{Computational Demands}
\label{sub:CPU}

\begin{figure}[h]
\begin{center}
\includegraphics[width=8.6 cm,angle=0,clip=true]{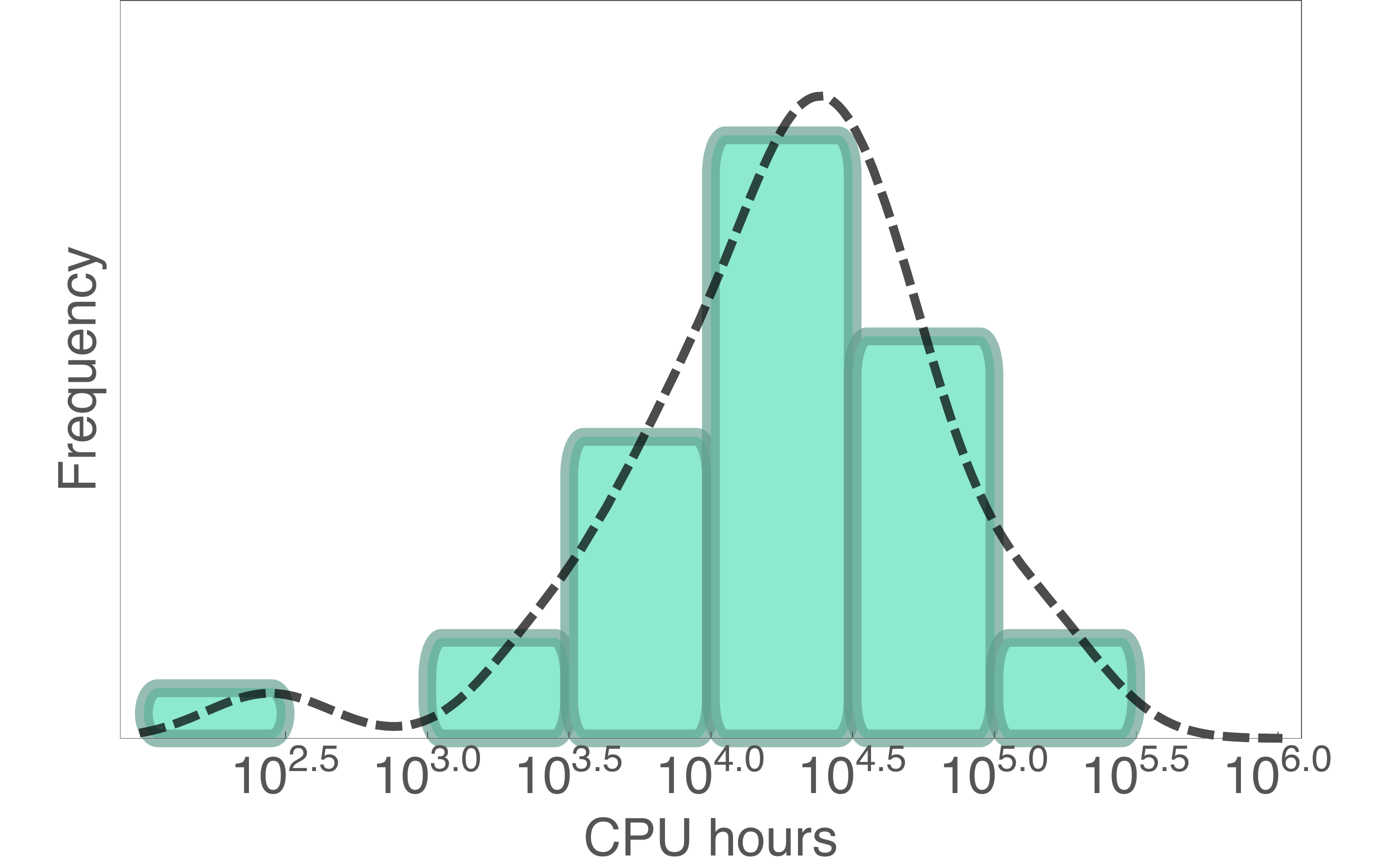}
\caption{
Histogram of the number of CPU hours used for model $\mathcal{H}_S$ of the 41 
KOIs surveyed in this work. The distribution has a mode of $24,000$\,CPU hours 
and a mean of $33,000$\,CPU hours.
} 
\label{fig:CPU_histo}
\end{center}
\end{figure}

We briefly discuss the computational demands of the HEK survey to date. Previous
surveys generally used less than the full \kepler\ time series (Q0-Q17) due to
the timeline of the data releases. In contrast, all 41 KOIs surveyed here used 
all available quarters, providing a fairer assessment of computational 
requirements for future surveys. We recorded the computational time used
for each model fit $\mathcal{H}_S$, in order to better understand our 
requirements. A histogram of the CPU times is shown in 
Figure~\ref{fig:CPU_histo}, with a mean time of 33,000\,hours per KOI. Note that 
these times do not include the planet-only fits (negligible times) and TTV fits
(sizable but less demanding). For all fits, the fits were conducted on AMD
Opteron 62xx/63xx CPUs. The spread in times appears mostly linked to the number
of transit epochs fitted, with larger time series naturally requiring more
overhead. With current in-house server capacity, we estimate that we are able to 
survey at least 100 KOIs per year.

\subsection{The Effects of Correlated Noise}
\label{sub:correlated}

Despite the groundbreaking precision achieved by \kepler\ 
\citep{christiansen:2012}, when one seeks low signal-to-noise events then the 
effects of time-correlated noise become important. This is particularly salient 
for exomoons, since the transits do not follow a simple ephemeris which can be
used to check for repeatability \citep{LUNA}. This allows the model great 
flexibility in explaining any slight random changes in flux, leading to
spurious detections (e.g. \citealt{KEP90}).

One strategy to guard against correlated noise is to use a likelihood function
more suited for such structure, for example using wavelets or Gaussian 
processes (pedagogical examples of these methods applied to photometry can be 
found in \citealt{carter:2009} and \citealt{gibson:2012}, respectively). 
However, these methods have so much flexibility than they may fit-out the
very low amplitude signals we seek. An alternative strategy, as employed in
this work, is to use vetting criteria, such as the double likelihood test F2
and insights from our photodynamical fits. Whilst maintaining the use of a 
Gaussian likelihood leads to more false-positives, it at least allows us see the 
full array of candidate signals and to assess them accordingly. Since the use
of Gaussian likelihood functions is so prevalent within the community, we here 
discuss the rate of false-positives induced by correlated noise structure.

\begin{figure}
\begin{center}
\includegraphics[width=8.8 cm,angle=0,clip=true]{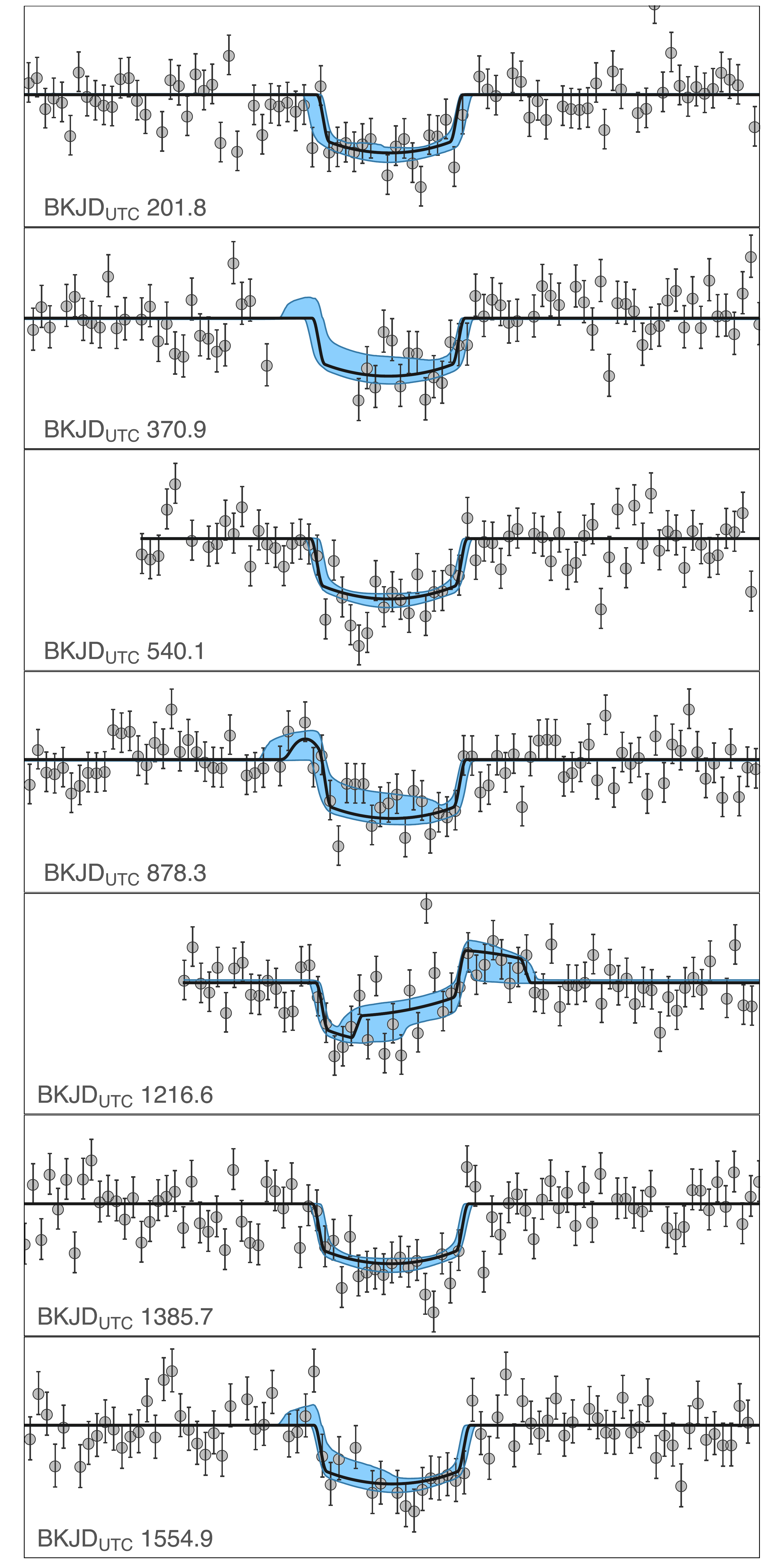}
\caption{
Example of a ``negative radius'' (inverted transits) exomoon solution, which 
occurs for around half of all KOIs surveyed. The transit light curves used for 
fitting KOI-4054.01 are shown, along with the maximum a-posteriori realization 
of our planet+moon model (black line) and the 95.45\% credible intervals (blue 
shaded region). This model is favored over a planet-only at 3.5\,$\sigma$ 
with $(R_S/R_P)=-0.71\pm0.09$, driven by correlated noise.
} 
\label{fig:KOI4054}
\end{center}
\end{figure}

Correlated noise directly affects the time series photometry, potentially
mimicking transit-like signals and leading to spurious detections of 
$(R_S/R_P)$. Whilst correlated noise can also affect the TTVs by distorting the 
planetary transit profile \citep{mazeh:2015}, such transit distortions are 
usually fitted out by the radius effect of our moon model, as seen in the 
example of Figure~\ref{fig:KOI4054}. In what follows, we focus on the effect of 
correlated noise on moon transit signals, i.e. parameter $(R_S/R_P)$. In our 
survey, a non-zero $(R_S/R_P)$ moon fit can be caused by two distinct reasons:

\begin{itemize}
\item[{\textbf{A]}}] The latent prior of $\rho_S<27.9$\,g\,cm$^{-3}$ (see
\S\ref{sec:analysis}) technically forbids $(R_S/R_P)=0$, thereby forcing
genuinely null $(R_S/R_P)$ cases to be either positive or negative.
\item[{\textbf{B]}}] Time correlated noise may drive the fits towards a 
false-positive likelihood minimum, since we adopt a Gaussian likelihood 
function.
\end{itemize}

It is of particular interest to understand how frequently case B occurs, which 
severely affects all methods seeking evidence of exomoons via their transits. In 
hindsight, not imposing the latent $\rho_S$ prior would be the most effective 
way to measure the occurrence of spurious $(R_S/R_P)$ detections due to 
correlated noise. However, we discuss here an alternative approach using the 
information in hand. The mean satellite density is computed using Equation~5 of 
\citet{weighing:2010}:

\begin{align}
\rho_S &= \rho_P \frac{ (M_S/M_P) }{ |R_S/R_P|^{3} }.
\end{align}

Since we impose the condition that $\rho_S<27.9$\,g\,cm$^{-3}$, then plotting
the joint posterior of $(\rho_P (M_S/M_P))$ and $|R_S/R_P|^3$ will have an
excluded region defined by a straight line passing through the origin of
gradient 27.9\,g\,cm$^{-3}$. We inspected each of the joint posteriors as
described, identifying cases where the posterior was bunched up against this
boundary, indicating the latent prior was likely driving a non-zero $(R_S/R_P)$
solution (i.e. case A). Of the 41 KOIs surveyed, we find that 22 show no
evidence for being affected by the latent $\rho_S$ prior and therefore the
non-zero $(R_S/R_P)$ solution is reasoned, by deduction, to be caused by time
correlated noise.

This estimate provides a handle as to the prevalence of correlated noise-induced
spurious moon-like transits, indicating that $(54\pm8)$\% of KOIs surveyed are 
affected. In future work, we intend to relax the latent priors and conduct a
more in-depth analysis of this question. Of the 22 identified cases, 8 yield 
negative radius moon solutions, compatible with the expectation of $11\pm2.3$, 
assuming that correlated noise induces negative radius transits with a $p$-value 
of 0.5.

If we had sought evidence for exomoons by seeking exomoon radius signals alone, 
and had not used a time-correlated noise likelihood, we would have erroneously 
inferred evidence for exomoons in around half of these cases, which is over one 
quarter ($\sim26$\%) of all KOIs surveyed. Thanks to photodynamics and our 
vetting criteria we were able to identify these spurious cases.

Although our survey strategy is robust against these time-correlated 
noise-induced moon-like transits, other exomoon search methods that do not use 
photodynamics may struggle to identify these frequent occurrences. We highlight 
these cases as a service to the community, to emphasize that exomoons live in 
the regime where correlated noise is present and one must employ methods to 
guard against it when seeking such signals.

\acknowledgements
\section*{Acknowledgements}

This work made use of the Michael Dodds Computing Facility, for which
we are grateful to Michael Dodds, 
Carl Allegretti, David Van Buren, Anthony Grange, Cameron Lehman, Ivan Longland, 
Dell Lunceford, Gregor Rothfuss, Matt Salzberg, Richard Sundvall, Graham 
Symmonds, Kenneth Takigawa,
Marion Adam, Dour High Arch, Mike Barrett, Greg Cole, Sheena Dean, Steven 
Delong, Robert Goodman, Mark Greene, Stephen Kitt, Robert Leyland, Matthias 
Meier, Roy Mitsuoka, David Nicholson, Nicole Papas, Steven Purcell, Austen 
Redman, Michael Sheldon, Ronald Sonenthal, Nicholas Steinbrecher, Corbin Sydney, 
John Vajgrt, Louise Valmoria, Hunter Williams, Troy Winarski and Nigel Wright.
DMK and DN acknowledge partial support 
from NASA grant NNX15AF09G. JH and GB acknowledge partial support from NASA 
grant NNX13AJ15G. JH acknowledges support from NNX14AF87G. DN acknowledges 
support from NSF AST-1008890. 
This paper includes data collected by the \kepler\ mission. Funding for the 
\kepler\ mission is provided by the NASA Science Mission directorate.
%


%


\begin{thebibliography}{99}
\bibitem[\protect\citeauthoryear{Ballard et al.}{2013}]{ballard:2013} 
Ballard, S., Charbonneau, D., Fressin, F., et al. 2013, ApJ, 773, 98
\bibitem[\protect\citeauthoryear{Barclay et al.}{2013}]{barclay:2013} 
Barclay, T., Rowe, J. F., Lissauer, J. J., et al. 2013, Nature, 494, 452
\bibitem[\protect\citeauthoryear{Barnes \& O'Brien}{2002}]{barnes:2002} 
Barnes, J. W. \& O'Brien, D. P. 2002, ApJ, 575, 1087
\bibitem[\protect\citeauthoryear{B\'eky et al.}{2014}]{beky:2014} 
B\'eky, B., Holman, M. J., Kipping, D. M. \& Noyes, R. W. 2014, ApJ, 788, 1
\bibitem[\protect\citeauthoryear{Borucki et al.}{2011}]{borucki:2011} 
Borucki, W. J., Koch, D. G., Basri, G., et al. 2011, ApJ, 736, 19
\bibitem[\protect\citeauthoryear{Borucki et al.}{2012}]{borucki:2012} 
Borucki, W. J., Koch, D. G., Batalha, N., et al. 2012, ApJ, 745, 120
\bibitem[\protect\citeauthoryear{Carter \& Winn}{2009}]{carter:2009}
Carter, J. A. \& Winn, J. N. 2009, ApJ, 704, 51
\bibitem[\protect\citeauthoryear{Christiansen et al.}{2012}]{christiansen:2012} 
Christiansen, J. L., Jenkins, J. M., Caldwell, D. A., et al. 2012, PASP, 124, 
1279
\bibitem[\protect\citeauthoryear{Claret \& Bloemen}{2011}]{claret:2011}
Claret, A. \& Bloemen, S. 2011, A\&A, 529, 75
\bibitem[\protect\citeauthoryear{Deck \& Agol}{2014}]{deck:2014} 
Deck, K. M. \& Agol, E. 2014, ApJ, submitted (arXiv:1411.0004)
\bibitem[\protect\citeauthoryear{Dumusque et al.}{2014}]{dumusque:2014} 
Dumusque, X., Bonomo, A. S., Haywood, R. D., et al. 2014, ApJ, 789, 154
\bibitem[\protect\citeauthoryear{Fabrycky et al.}{2014}]{fabrycky:2014} 
Fabrycky, D. C., Lissauer, J. J., Ragozzine, D., et al. 2014, ApJ, 790, 146
\bibitem[\protect\citeauthoryear{Feroz et al.}{2008}]{feroz:2008} 
Feroz, F. \& Hobson, M. P. 2008, MNRAS, 384, 449
\bibitem[\protect\citeauthoryear{Feroz et al.}{2009}]{feroz:2009} 
Feroz, F., Hobson, M. P. \& Bridges, M. 2009, MNRAS, 398, 1601
\bibitem[\protect\citeauthoryear{Feroz et al.}{2013}]{feroz:2013}
Feroz, F. 2013, ``Proceedings 59th ISI World Statistics Congress'', 
25-30 August 2013, Hong Kong, p. 2081
\bibitem[\protect\citeauthoryear{Fewell}{2006}]{fewell:2006} 
Fewell, M. 2006, ``Area of Common Overlap of Three Circles'', Tech. Rep. 
DSTO-TN-0722 [http://hdl.handle.net/1947/4551]
\bibitem[\protect\citeauthoryear{Fischer et al.}{2012}]{fischer:2012} 
Fischer, D. A., Schwamb, M. E., Schawinski, K., et al. 2012, MNRAS, 419, 2900
\bibitem[\protect\citeauthoryear{Gaudi}{2006}]{gaudi:2006} 
Gaudi, B. S. 2006, arXiv:astro-ph/0612141
\bibitem[\protect\citeauthoryear{Gibson et al.}{2012}]{gibson:2012}
Gibson, N. P., Aigrain, S., Roberts, S., et al. 2012, MRNAS, 419, 2683
\bibitem[\protect\citeauthoryear{Handley et al.}{2015}]{handley:2015}
Handley, W. J., Hobson, M. P. \& Lasenby, A. N. 2015, MNRAS, submitted 
(arXiv:1502.01856)
\bibitem[\protect\citeauthoryear{Heller et al.}{2014}]{OSE:2014}
Heller, R. 2014, ApJ, 787, 14
\bibitem[\protect\citeauthoryear{Huang et al.}{2013}]{huang:2013} 
Huang, X., Bakos, G. A. \& Hartman, J. D. 2013, MNRAS, 429, 2001
\bibitem[\protect\citeauthoryear{Kipping}{2009a}]{kipping:2009} 
Kipping, D. M. 2009a, MNRAS, 392, 181
\bibitem[\protect\citeauthoryear{Kipping}{2009b}]{kipping:2009b} 
---. 2009b, MNRAS, 396, 1793
\bibitem[\protect\citeauthoryear{Kipping}{2010a}]{binning:2010} 
---. 2010a, MNRAS, 408, 1758
\bibitem[\protect\citeauthoryear{Kipping}{2010b}]{weighing:2010} 
---. 2010b, MNRAS, 409, L119
\bibitem[\protect\citeauthoryear{Kipping}{2011}]{LUNA} 
---. 2011, MNRAS, 416, 689
\bibitem[\protect\citeauthoryear{Kipping}{2013}]{LDfitting:2013} 
---. 2013, MNRAS, 435, 2152
\bibitem[\protect\citeauthoryear{Kipping et al.}{2009}]{kippingetal:2009} 
Kipping, D. M., Fossey, S. J. \& Campanella, G. 2009, MNRAS, 400, 398
\bibitem[\protect\citeauthoryear{Kipping \& Tinetti}{2010}]{nightside:2010} 
Kipping, D. M. \& Tinetti, G. 2010, MNRAS, 407, 2589
\bibitem[\protect\citeauthoryear{Kipping et al.}{2012}]{HEK1} 
Kipping, D. M., Bakos, G. \'A., Buchhave, L., Nesvorn\'y, D. \& Schmitt, A.
2012, ApJ, 750, 115
\bibitem[\protect\citeauthoryear{Kipping et al.}{2013a}]{HEK2} 
Kipping, D. M., Hartman, J., Buchhave, L. A., et al. 2013a, ApJ, 770, 101
\bibitem[\protect\citeauthoryear{Kipping et al.}{2013b}]{HEK3} 
Kipping, D. M., Forgan, D., Hartman, J., et al. 2013b, ApJ, 777, 134
\bibitem[\protect\citeauthoryear{Kipping et al.}{2014a}]{HEK4} 
Kipping, D. M., Nesvorn\'y, D., Buchhave, L. A., et al. 2014a, ApJ, 784, 28
\bibitem[\protect\citeauthoryear{Kipping et al.}{2014b}]{KEP421}
Kipping, D. M., Torres, G., Buchhave, L. A., et al. 2014b, ApJ, 795, 25
\bibitem[\protect\citeauthoryear{Kipping et al.}{2015}]{KEP90}
Kipping, D. M., Huang, X., Nesvorn\'y, D., et al. 2015, ApJ, 799, 14
\bibitem[\protect\citeauthoryear{Kundurthy et al.}{2011}]{kundurthy:2011} 
Kundurthy, P., Agol, E., Becker, A. C., et al. 2011, ApJ, 731, 123
\bibitem[\protect\citeauthoryear{Lewis}{2014}]{lewis:2014} 
Lewis, K. 2014, STScI ``Habitable Worlds Across Time and Space'' meeting,
oral presentation
\bibitem[\protect\citeauthoryear{Lucy \& Sweeney}{1971}]{lucy:1971}
Lucy, L. B. \& Sweeney, M. A. 1971, 76, 544
\bibitem[\protect\citeauthoryear{Mandel \& Agol}{2002}]{mandel:2002} Mandel, K. 
\& Agol, E. 2002, ApJ, 580, 171
\bibitem[\protect\citeauthoryear{Mazeh et al.}{2013}]{mazeh:2013} 
Mazeh, T., Nachmani, G., Holczer, T., et al. 2013, ApJS, 208, 16
\bibitem[\protect\citeauthoryear{Mazeh et al.}{2015}]{mazeh:2015} 
Mazeh, T., Holczer, T. \& Shporer, A. 2015, ApJ, 800, 142
\bibitem[\protect\citeauthoryear{Namouni}{2010}]{namouni:2010} 
Namouni, F. 2010, ApJ, 719, 145
\bibitem[\protect\citeauthoryear{Nesvorn\'y et al.}{2012}]{KOI872} 
Nesvorn\'y, D., Kipping, D. M., Buchhave, L., et al. 2012, Science, 336, 1133
\bibitem[\protect\citeauthoryear{Nesvorn\'y \& Vokrouhlick\'y}{2014}]{nesvorny:2014} 
Nesvorn\'y, D. \& Vokrouhlick\'y, D. 2014, ApJ, 790, 58
\bibitem[\protect\citeauthoryear{Raymond et al.}{2007}]{raymond:2007} 
Raymond, S. N., Scalo, J. \& Meadows, V. S. 2007, ApJ, 669, 606
\bibitem[\protect\citeauthoryear{Rowe et al.}{2014}]{rowe:2014} 
Rowe, J. F., Bryson, S. T., Marcy, G. W., et al. 2014, ApJ, 784, 45
\bibitem[\protect\citeauthoryear{Sartoretti \& Schneider}{1999}]{sartoretti:1999}
Sartoretti, P. \& Schneider, J. 1999, A\&AS, 134, 553
\bibitem[\protect\citeauthoryear{Schwamb et al.}{2014}]{schwamb:2014} 
Schwamb, M. E., Orosz, J. A., Carter, J. A., et al. 2013, ApJ, 768, 127
\bibitem[\protect\citeauthoryear{Simon et al.}{2009}]{simon:2009}
Simon, A. E., Szab\'o, Gy. M. \& Szatm\'ary, K. 2009, EM\&P, 105, 385
\bibitem[\protect\citeauthoryear{Simon et al.}{2012}]{simon:2012}
Simon, A. E., Szab\'o, Gy. M., Kiss, L. L. \& Szatm\'ary, K. 2012, MNRAS,
419, 164
\bibitem[\protect\citeauthoryear{Skilling}{2004}]{skilling:2004} 
Skilling, J. 2004, in Fischer R., Preuss R., Toussaint U. V., eds,
American Institute of Physics Conference Series Nested Sampling. pp 395–405
\bibitem[\protect\citeauthoryear{Smith et al.}{2012}]{smith:2012} 
Smith, J. C., Stumpe, M. C., Van Cleve, J. E., et al. 2012, PASP, 124, 1000
\bibitem[\protect\citeauthoryear{Stumpe et al.}{2012}]{stumpe:2012} 
Stumpe, M. C., Smith, J. C., Van Cleve, J. E., et al. 2012, PASP, 124, 985
\bibitem[\protect\citeauthoryear{Szab\'o et al.}{2013}]{szabo:2013}
Szab\'o, R., Szab\'o, Gy. M., D\'alya, G., et al. 2013, A\&A, 553, 17
\bibitem[\protect\citeauthoryear{Torres et al.}{2015}]{torres:2015} 
Torres, G., Kipping, D. M., Fressin, F., et al. 2015, ApJ, 800, 99
\bibitem[\protect\citeauthoryear{Wang et al.}{2013}]{wang:2013} 
Wang, J., Fischer, D. A., Barclay, T., et al. 2013, ApJ, 776, 10
\bibitem[\protect\citeauthoryear{Zsom et al.}{2013}]{zsom:2013} 
Zsom, A., Seager, S., de Wit, J. \& Stamenkovi\'c, V. 2013, ApJ, 778, 109
\end{thebibliography}
\end{document}